\documentclass[journal,12pt,onecolumn]{IEEEtran}

\ifCLASSINFOpdf
\else
\fi
\usepackage[explicit]{titlesec}
\usepackage{graphicx}
\usepackage{float}
\usepackage{epstopdf}
\usepackage{subfigure} 
\usepackage{algorithmic}
\usepackage{setspace}
\usepackage{amsmath}
\usepackage{amsthm}
\usepackage{cite}
\newtheorem{theorem}{Theorem}

\usepackage{amssymb}
\newtheorem{remark}{Remark}

\newcommand{\RNum}[1]{\uppercase\expandafter{\romannumeral #1\relax}}
\newtheorem{lemma}{Lemma}

\usepackage{stfloats}
\newtheorem{problem}{Problem}
\usepackage{color}
\usepackage{tikz,xcolor,hyperref}
\definecolor{lime}{HTML}{A6CE39}%
\titlespacing{\section}{0pt}{0ex plus .0ex minus .0ex}{.3ex plus .0ex}
\titlespacing{\subsection}{0pt}{0ex plus .0ex minus .0ex}{.3ex plus .0ex}

\makeatletter
\DeclareRobustCommand{\orcidicon}{%
	\begin{tikzpicture}
	\draw[lime, fill=lime] (0,0) 
	circle [radius=0.16] 
	node[white] {{\fontfamily{qag}\selectfont \tiny ID}};    \draw[white, fill=white] (-0.0625,0.095) 
	circle [radius=0.007];    \end{tikzpicture}
	\hspace{-2mm}}
\foreach \x in {A, ..., Z}{%
\expandafter\xdef\csname orcid\x\endcsname{\noexpand\href{https://orcid.org/\csname orcidauthor\x\endcsname}{\noexpand\orcidicon}}
}

\newcommand*\bigcdot{\mathpalette\bigcdot@{.5}}
\newcommand*\bigcdot@[2]{\mathbin{\vcenter{\hbox{\scalebox{#2}{$\m@th#1\bullet$}}}}}
\makeatother

\usepackage[linesnumbered,ruled,vlined]{algorithm2e}

\begin{document}
%
\title{New Upper Bounds on the Error Probability under ML Decoding for Spinal Codes and the Joint Transmission-Decoding System Design}

\author{Aimin Li\orcidA{}, 
		\emph{Graduate Student Member, IEEE,}
	Shaohua Wu\orcidB{}, 
		\emph{Member, IEEE,}
	Jian Jiao\orcidD{}, 
		\emph{Member, IEEE,}
	Ning Zhang\orcidE{}, 
		\emph{Senior Member, IEEE,}
	and Qinyu Zhang\orcidF{},
		\emph{Senior Member, IEEE.}
	\thanks{
			A. Li, S. Wu, J. Jiao and Q. Zhang are with the Department of Electronic Engineering, Harbin Institute of Technology (Shenzhen), Guangdong, China. N. Zhang is with the Department of Electrical and Computer Engineering, University of Windsor, Windsor, ON, N9B 3P4, Canada (e-mail: liaimin@stu.hit.edu.cn; hitwush@hit.edu.cn; jiaojian@hit.edu.cn; ning.zhang@uwindsor.ca; zqy@hit.edu.cn).
	}
}

\maketitle

\begin{abstract}
Spinal codes are a type of capacity-achieving rateless codes that have been proved to approach the Shannon capacity over the additive white Gaussian noise (AWGN) channel and the binary symmetric channel (BSC). In this paper, we aim to analyze the bounds on the error probability of Spinal codes and design a joint transmission-decoding system. First, in the finite block-length regime, we derive new upper bounds on the Maximum Likelihood (ML) decoding error probability for Spinal codes over both the AWGN channel and the BSC. Then, based on the derived bounds, we formulate a rate maximization problem. As the solution exhibits an incremental-tail-transmission pattern, we propose an improved transmission scheme, referred to as the thresholded incremental tail transmission (TITT) scheme. Moreover, we also develop a dynamic TITT-matching decoding algorithm, called the bubble decoding with memory (BD-M) algorithm, to reduce the decoding time complexity. The TITT scheme at the transmitter and the BD-M algorithm at the receiver jointly constitute a dynamic transmission-decoding system for Spinal code, improving its rate performance and decoding throughput. Theoretical analysis and simulation results are provided to verify the superiority of the derived bounds and the proposed joint transmission-decoding system design.
\end{abstract}


\IEEEpeerreviewmaketitle

\section{Introduction}
\subsection{Background}
\IEEEPARstart{F}{irst} proposed by Perry \emph{et al.} in \cite{ref1}, Spinal codes are a new type of near-Shannon-capacity channel coding techniques that have been theoretically proved to achieve the Shannon capacity over both the additive white Gaussian noise (AWGN) channel and the binary symmetric channel (BSC) \cite{ref2}. The core idea in this code is applying a sequential of hash functions combing with the random number generators (RNGs) to generate pseudo-random coded symbols: the hash functions provide the coded symbols with the pairwise independent feature, which thereby ensures good resilience to noise and errors; the RNGs inherit from the idea of Shannon random coding, which can generate pseudo-random coded symbols as much as possible to cope with the time-varying channel by nature. Spinal code draws its strength from the excellent code rate performance. It has been shown in \cite{Spinalcodes} that rateless Spinal codes exhibit superiority in rate performance when compared to fixed-rate LDPC codes \cite{LDPCcodes}, rateless Raptor codes \cite{Raptorcodes}, and the layered rateless coding approach \cite{ratelesscodes} of Strider codes \cite{stridercodes}, across various channel conditions and message sizes.

In addition to the  capacity-approaching code rate performance, a more attractive feature for Spinal codes is that Spinal codes are also a family of \emph{rateless codes}. Different from conventional fixed-rate codes that should dynamically pre-define the ``bit rate'', i.e., a modulation mode, channel code, and code rate,   Spinal codes transmit coded symbols in a rateless mode. By rateless transmission, Spinal codes do not need to frequently estimate the time-varying channel state to dynamically adjust the code rate and the modulation mode. Instead, they keep generating and transmitting pseudo-random coded symbols until the transmitted symbols are sufficient for the decoder to decode successfully and an acknowledgment (ACK) is fed back to the transmitter to interrupt the transmission. By this means, Spinal codes themselves are a type of advanced adaptive coding modulation technique, which enables adaptive parameter selection to ensure reliable and efficient end-to-end communication.

\subsection{Literature Review}

Attributed to the near-Shannon-capacity characteristic and the rateless feature, Spinal codes have aroused extensive research interest in both performance optimization and theoretical analysis. One of the most popular and open challenges in facilitating Spinal codes from theory to practice is to design a practical low-complexity decoding algorithm for Spinal codes. Formally, such a challenge exists due to the high-frequency use of hash functions and the multiple rounds of tentative decoding. To circumvent this problem, traditionally, the tree pruning strategies have been extensively explored to design the low-complexity decoding algorithms \cite{ref1,ref6,ref9,ref10}. In \cite{Spinalcodes}, the proposed bubble decoding algorithm is the initial practical decoding algorithm customized for Spinal codes. Then, inspired by the tree-structure feature of Spinal codes, Yang \emph{et al.} introduce the forward stack decoding (FSD) algorithm, which was first proposed in \cite{ref8},  to implement the decoding process of Spinal codes \cite{ref6}. By such an implementation, the FSD algorithm showcases much lower decoding time complexity compared to the bubble decoding algorithm without sacrificing rate performance. In \cite{ref9}, the proposed sliding window decoding (SFD) algorithm further reap benefits in much lower decoding time complexity compared to the FSD algorithm. A trade-off between decoding efficiency and decoding reliability is also investigated. In \cite{ref10}, Hu \emph{et al.} introduces several dynamic parameters in the decoding process Spinal codes. The proposed algorithm, referred to as the block dynamic decoding (BBD) algorithm, also demonstrates superior performance in both the decoding complexity and the decoding error probability. Nevertheless, we notice that the above works mainly focus on designing the pruning strategy to eliminate the calculation required in a single decoding round. For rateless Spinal codes, the transmission and decoding process are both consecutive and not unique. In this regard, we would like to pay attention to investigating the dynamic changes of the decoding tree between adjacent decoding rounds of rateless Spinal codes.

Puncturing on the coded symbols is a commonly adopted transmission design to improve the rate performance of channel coding techniques. In \cite{ref11}, coding schemes that can be punctured to adapt to time-varying channels are known as rate-compatible codes. As the redundant symbols are punctured, the code rate increases by nature. As such, the puncturing technique has drawn extensive interest in the literature, such as punctured LDPC codes \cite{ref12,ref3,ref4}, punctured Turbo codes \cite{Turbocodes}, and punctured Polar codes \cite{Polarcodes,Polarcodes3}. For rateless Spinal codes, uniform puncturing is one of the most commonly used puncturing schemes, where the coarse-grained transmission unit, pass, is refined into some finer-grained sub-passes, and the transmitter only transmits a single symbol in a sub-pass \cite{Spinalcodes}. In \cite{ref15}, uniform puncturing is also adopted in the proposed S-Spinal codes to achieve finer granularity for compression purposes. However, the uniform puncturing implementation is only a general-purpose puncturing method that has been applied extensively in rateless coded systems\cite{ref16}. In such a case, the puncturing scheme that customizes the unique properties of Spinal codes is expected to be designed. 

Performance evaluation of channel codes is another important work in understanding the essence of a coding technique. In addition to Monte Carlo simulations, which require amounts of repeated calculations, the bounding technique is a more straightforward method to evaluate the performance of a coding technique\footnote{In most of the cases, to derive the exact closed-form expression to evaluate the performance of a near-Shannon-limit coding technique is intractable, and thus to derive the bound is an alternative.}. The recent focus given to the performance bounds of codes is on some near-Shannon-limit performing codes, ranging from the performance bounds on LDPC codes \cite{LDPCbound,LDPCbound2}, to other advanced tight bounds on Turbo codes\cite{Trubobound1,Trubobound2,Trubobound3}, and Polar codes\cite{Polarbound1,Polarbound2}. However, for Spinal codes, which are a new family of near-Shannon-capacity codes, the theoretical performance analysis is still in its infancy. One of the first works devoted to the performance of Spinal codes is that of Yu \emph{et al.}\cite{ref18}, wherein the tree structure of Spinal codes is considered, and some general bounding techniques on pairwise independent random codes, i.e., the Random Coding Union (RCU) bound and a variant of Gallager's result, are introduced as cores to obtain the closed-form upper bounds on the error probability of Spinal codes. Nevertheless, we notice that derived upper bounds are not customized for Spinal codes, but play as applications of the RCU bound and Gallager's result. In this regard, the bounds on the error probability exhibit loose performance (see Section \ref{section6}). In our previous work \cite{ref19}, we analyze the upper bound on the error probability of Spinal codes over the BSC. However, the derived upper bound in \cite{ref19} remains several static parameters obtained through simulation implementations. Therefore, the bounds in \cite{ref19} and are not completely explicit. To this end, we would like to derive explicit upper bounds on the error probability of Spinal codes. Instead of taking the general upper bounds of pairwise independent random codes as cores, we consider the coding, decoding and transmission process of Spinal codes in detail to derive the upper bounds on the error probability for Spinal codes in the FBL regime.

\subsection{Contributions}
Motivated by the above, the main contributions of this work are summarized as follows. We begin with deriving the upper bounds on the error probability of ML-decoded Spinal codes over both the BSC and the AWGN channel under the finite block-length (FBL) regime. The derivation leads to the closed-form upper bounds shown in Section \ref{section3}, and demonstrate tighter performance at high SNRs as shown in Section \ref{section6}. Then, based on the derived closed-form upper bounds, we formulate a rate maximization problem and correspondingly design a customized iterative algorithm to solve it. Notice that the optimal solution to this problem forms an incremental-tail-transmission pattern; heuristically, we propose the thresholded incremental tail transmission (TITT) scheme for Spinal codes. Furthermore, to address the high decoding complexity issue at the receiver, we investigate the dynamic changes of the decoding tree between adjacent decoding tentatives and show that only parts of the decoding tree change during the consecutive decoding process, and thus the tree reconstruction of the unchanged parts can be skipped. To this end, we find that most of the available decoding algorithms for Spinal codes in the literature can be improved by simply introducing an extra memory to restore the unchanged parts of the decoding tree. For the experimental purpose, we take the most commonly used bubble decoding algorithm as an example, and design a bubble-decoding-based dynamic decoding algorithm, referred to as the bubble decoding with memory (BD-M). By combing the proposed TITT scheme at the transmitter with the BD-M algorithm at the receiver, we further observe that the BD-M algorithm can match perfectly with the tail-incremental-transmission property of the TITT scheme, enjoying both higher rate performance and lower decoding complexity.

\subsection{Organization}
The rest of this paper is organized as follows. Section \ref{section2} introduces the preliminaries of Spinal codes. In Section \ref{section3}, the new upper bounds on the error probability of ML-decoded Spinal codes are derived. The details of the transmission scheme design are presented in Section \ref{section4}. In Section \ref{section5}, the details of the BD-M algorithm and its complexity analysis are presented. In Section \ref{section6}, simulation results are provided, followed by conclusions in Section \ref{section7}.
\section{Preliminaries}

\label{section2}
\subsection{Encoding Process of Spinal Codes}
\begin{figure}[htbp]
	\centering
	\includegraphics[angle=0,width=0.8\textwidth]{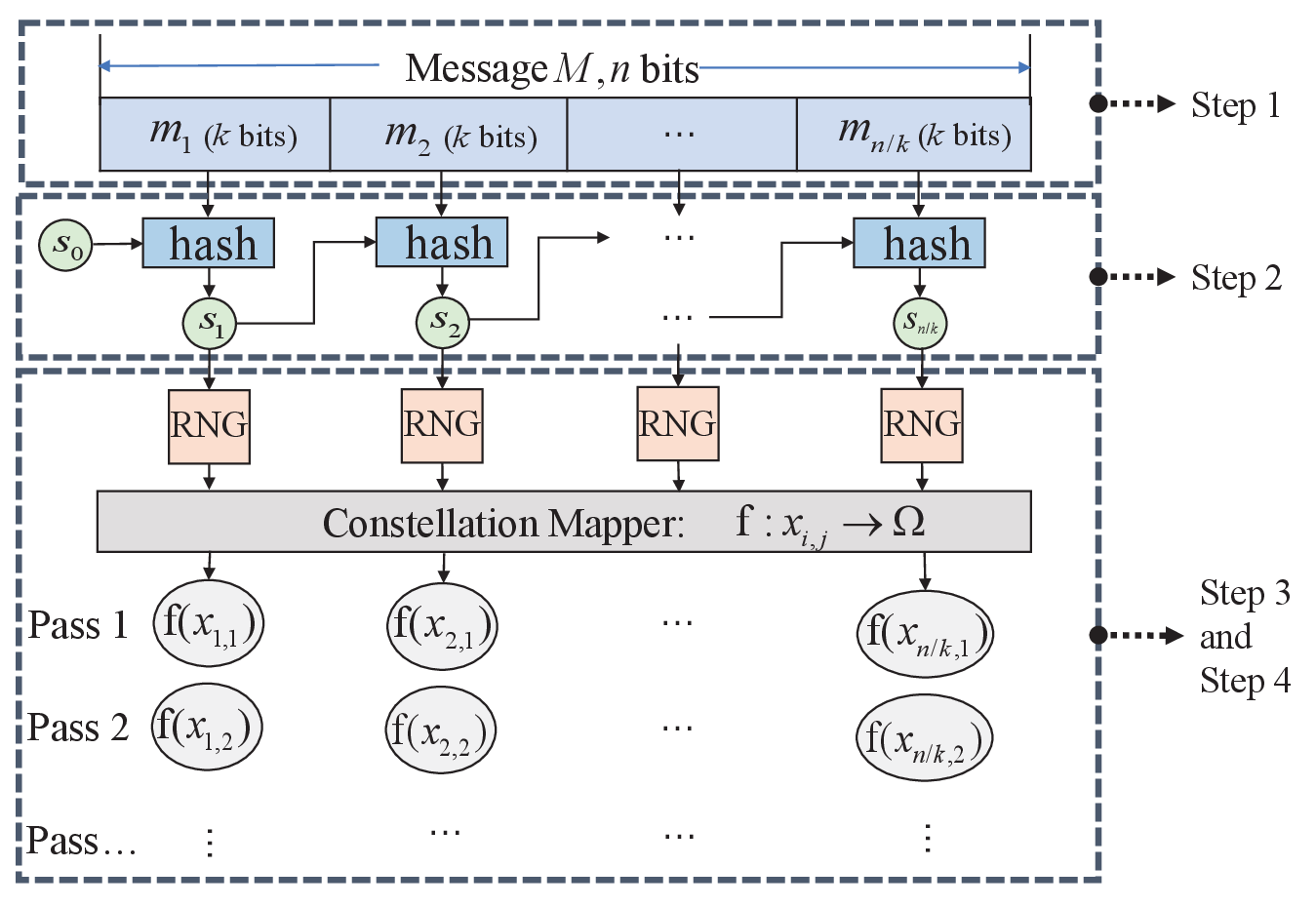}
	\caption{The encoding process of Spinal codes.}\label{encoding process of Spinal codes}
\end{figure}
 As shown in Fig. \ref{encoding process of Spinal codes}, the encoding process of Spinal codes can be accomplished in four steps:
\begin{enumerate}
	\item An $n$-bit message $M$ is divided into $k$-bit segments, denoted by $m_i$, where $i=1, 2, \ldots, n/k$.	
	\item An iterative process is invoked to generate the series of $v$-bit spine values ${s_i}$: 
	\begin{equation}\label{spine}
	{s_i} = {\rm{ }}h({s_{i - 1}},{\rm{ }}{m_i}), i=1, \ldots, n/k, s_0=0^{v}. \footnote{The initial spine value $s_0$ is known to both the encoder and the decoder. Without loss of generality, we set $s_0=0^v$ in this paper.}
	\end{equation}
	\item The $v$-bit spine value ${s_i}$ serves as a seed of an RNG to generate a pseudo-random uniform-distributed sequence $\left\{{x_{i,j}}\right\}$:
	\begin{equation}
	{\rm{RNG: }}{s_i} \mathop  \to \left\{ {x_{i,j}} \right\}, j = 1,2,3,\cdots,
	\end{equation}
	where ${x_{i,j}}\in{\left\{ {0{\rm{,}}1} \right\}^c}$, ${s_i}\in{\left\{ {0{\rm{,}}1} \right\}^v}$.
	\item The constellation mapper maps each c-bit symbol to a channel input set $\Omega$:
	\begin{equation}
	{\operatorname f:{x_{i,j}}\mathop  \to \Omega,}
	\end{equation}
	where $\operatorname{f}$ is a constellation mapping function and ${\Omega}$ denotes the channel input set. In this paper, we consider the uniform constellation mapper, where $\operatorname{f}$ converts each $c$-bit symbol $x_{i,j}$ to a decimal symbol $\operatorname{f}\left(x_{i,j}\right)$. 
\end{enumerate}
\begin{remark}\label{PIP}
	{\rm As Perry \emph{et al.} indicates, the hash function employed by Spinal codes should have pairwise independent property: \cite{Spinalcodes} }
	\begin{equation} 
	\begin{array}{l}
	\mathbb{P}\left( h\left( s,m\right)=x, h\left(  s^\prime,m^\prime\right)=x^\prime \right) \\
	=\mathbb{P}\left( h\left( s,m\right)=x \right)
	\cdot\mathbb{P}\left( h\left( s^\prime,m^\prime\right)=x^\prime \right)\\
	=2^{-2v},
	\end{array}
	\end{equation}
	{\rm where $\left( s,m \right)$ and $\left( s^\prime,m^\prime \right)$ are two different inputs of the hash function.}	
\end{remark}

\subsection{Rateless Transmission}

 The original transmission scheme for Spinal codes is a pass-by-pass transmission scheme. Denote ${\textbf{X}_{j}} = [\operatorname{f}\left(x_{1,j}\right),\operatorname{f}\left(x_{2,j}\right), \ldots ,\operatorname{f}\left(x_{n/k,j}\right)]$ as the $j^{\rm th}$ pass of coded symbols. The conventional pass-by-pass transmission scheme will form a transmission process of ${\textbf{X}_{j}} \to {\textbf{X}_{j+1}} \to {\textbf{X}_{j+2}} \to \cdots$. This transmission process ends only when the number of received symbols is enough for successful decoding and an ACK is sent to interrupt the transmission. Note that correct decoding may be completed in any round of transmission, the cumulative transmitted symbols before correct decoding must be a multiple of $n/k$, belonging to the set 
$\left\{in/k |i\in \mathbb{Z}^+ \right\}$.

In \cite{Spinalcodes}, the authors propose a uniform-puncturing-based transmission scheme for Spinal codes. Instead of transmitting Spinal codes pass by pass, the uniform-puncturing-based transmission scheme transmits Spinal codes in a symbol by symbol pattern. By this means, the cumulative transmitted symbols before correct decoding belongs to a finer set $\left\{i |i\in \mathbb{Z}^+ \right\}$. Take a typical $(n,k)$ Spinal codes as an example. Let ${\bf g} = [g_{1},g_{2},\ldots,
g_{n/k}]$ denote the transmission order of the symbols. The uniform-puncturing-based transmission scheme forms a transmission process of $\operatorname{f}\left(x_{g_1,j}\right) \to \operatorname{f}\left(x_{g_2,j}\right)  \to \cdots \to \operatorname{f}\left(x_{g_{n/k},j}\right) \to \operatorname{f}\left(x_{g_1,j+1}\right) \to \operatorname{f}\left(x_{g_2,j+1}\right) \to \cdots$. The transmitter will continue the transmission process until an ACK is received.

For the uniform-puncturing-based transmission scheme, the punctured redundant symbols are from the last pass of transmission. For example, assuming that only $in/k+1$ symbols need to be sent for successful decoding of uniform-puncturing-based transmission scheme, then under the same condition, the traditional pass-by-pass transmission scheme needs to send additional $n/k-1$ symbols in the ${i+1}$ pass to decode correctly since its transmission unit is pass. Compared with the traditional pass-by-pass transmission scheme, the uniform-puncturing-based transmission scheme punctured out $n/k-1$ symbols, thus improving the rate performance of Spinal codes. To conclude, the uniform-puncturing-based transmission scheme reduces the transmission of redundant symbols in the last pass and improves the code rate of Spinal codes by refining the unit of each transmission from the coarse-grained pass to the finer-grained symbol.

Although the uniform-puncturing-based transmission scheme indeed improves the rate performance of Spinal codes, there are still some shortcomings that can be improved. First, the uniform-puncturing-based transmission scheme needs to completely transmit a whole pass of symbols before continuing to transmit the subsequent pass, which limits its puncturing coverage to the last transmitted pass only. A more flexible transmission scheme without this limitation has yet to be proposed. Second, the symbol-by-symbol transmission pattern increases the decoding attempts at the receiver and thus reduces the decoding throughput of Spinal codes. To resolve this issue, we design a dynamic decoding framework for Spinal codes. The proposed decoder can dynamically store the decoding information in each decoding process and utilize it to assist subsequent decoding.

\subsection{Decoding Process of Spinal Codes}
\label{section2c}
For Spinal codes over the AWGN channel, the optimal decoding algorithm is maximum likelihood (ML) decoding. In ML decoding, the decoder adopts the shared knowledge of the same hash function, the same initial spine value $s_{0}$ and the same RNG to replay the coding process over the set of received symbols. By adopting the shared knowledge, the decoder aims to find the best matching sequence $\hat{M} \in {\left\{ {0,1} \right\}^n}$ whose coded vector $ \textbf{x}(\hat M)$ is closest to the received vector $\textbf{y}$ in Euclidean distance. 

 For conventional pass-by-pass transmission scheme, let $L$ denote the number of transmitted passes for Spinal codes, the ML decoding rule can be expressed as:
\begin{equation} 
 \hat{M}= \mathop{\arg\min}_{M'\in\left\{0,1\right\}^n}\| \textbf{y}-\textbf{x}(M')\|^2=\mathop{\arg\min}_{M'\in\left\{0,1\right\}^n}\sum_{i=1}^{n/k}\sum_{j=1}^{L}({y}_{i,j}-\operatorname{f}\left({x}_{i,j}({M}')\right))^2,
\end{equation}
where $\hat M$ is the decoding result, $M'$ represents the candidate sequence.

 For puncturing symbol-by-symbol transmission scheme, let $\ell_{i}$ denote the number of transmitted symbols generated from the $i^{\rm th}$ spine value $s_i$, the ML decoding rule should be slightly 
\begin{equation}\label{eq1} 
\hat{M}=\mathop{\arg\min}_{M'\in\left\{0,1\right\}^n}\| \textbf{y}-\textbf{x}(M')\|^2\\
=\mathop{\arg\min}_{M'\in\left\{0,1\right\}^n}\sum_{i=1}^{n/k}\sum_{j=1}^{\ell_{i}}({y}_{i,j}-\operatorname{f}\left({x}_{i,j}({M}'\right)))^2.
\end{equation}
For the BSC, the only difference is that the Euclidean distance should be replaced with Hamming distance.

However, traversing all candidate sequences $M'$ results in an exponential increase in complexity. The serial coding structure of Spinal codes determines that Spinal codes are a type of tree code. As a result, the bubble decoding algorithm proposed in \cite{Spinalcodes} can be applied to decrease the decoding complexity. Instead of searching the entire decoding tree, a bubble decoder calculates the cost $C_\lambda$ for the nodes at each layer and sorts them to retain only the $B$ best candidate nodes with the lowest cost at each layer. The reserved nodes are then expanded to $B2^{k}$ child nodes. This repeats until the bubble decoder ends up with a list of $B$ messages at the last layer, from which the best one is selected as the decoding output. The cost at each layer can be calculated as:
  \begin{equation}\label{layer} 
  \begin{split}
  C_\lambda=\sum_{i=1}^{\lambda}\sum_{j=1}^{\ell_{i}}\|{y}_{i,j}-\operatorname{f}({x}_{i,j}({M}'))\|^2,
  \end{split}
  \end{equation}
where $\lambda$ denotes the currently expanded number of layers.

\section{Upper bounds on the error Probability} \label{section3}
\subsection{Upper Bounds Based on General Bounding Techniques}
Here we simply review the upper bounds obtained in \cite{ref18}, where the authors have derived the upper bounds on the error probability of ML decoding for Spinal codes over the AWGN channel and the BSC, respectively.
\subsubsection{The Bound Based on Gallager's Result}
\begin{theorem} \label{lemma1}
	{\rm \textbf{(The upper bound on the error probability over the AWGN channel \cite{ref18})} Consider Spinal codes with message length $n$ and segmentation parameter $k$ transmitted over an AWGN channel with noise variance $\sigma^{2}$. Let $L$ be the number of transmitted passes. The average error probability under ML decoding can be upper bounded by}
	\begin{equation} \label{oldAbound}
	P_{e} \leq 1 - \prod_{a=1}^{n/k}(1-\epsilon_{A}(L_{a}, U_{a})),
	\end{equation}
	{\rm with}
	\begin{equation} \label{Gallagerbound}
	\epsilon_{A}(L_{a}, U_{a}) = 2^{-L_{a}(E_{o}(\mathcal{Q})-\frac{\log U_{a}}{L_{a}})},\tag{\ref{oldAbound}a}
	\end{equation} 
	{\rm where $L_{a}=L(n/k-a+1)$, $U_{a}=2^{k(n/k-a+1)}$, $\mathcal{Q}$ denotes the probability distribution of the channel input and
	\begin{equation} \label{E0Q}
		E_{0}\left(\mathcal{Q}\right)=-\log_{2}\left\lbrace\frac{1}{\sqrt{2\pi\sigma^{2}}}\times \int_{\mathbb{R}}{\left(\sum_{x \in\Omega}\mathcal{Q}\left(x \right) \cdot\exp \left( -\frac{\left(y-x\right) ^{2}}{4\sigma^{2}} \right)\right)^{2}dy} \right\rbrace.\tag{\ref{oldAbound}b}
		\end{equation}
	}
\end{theorem}
The core of Theorem \ref{lemma1} is to apply a general bounding technique developed by Gallager's result in \cite{ref17}. That is ${P_e} \le {2^{ -L_a\left( {{E_0}\left( \mathcal{Q} \right) - R} \right)}}$, where $R=n/L_a$ represents the code rate. In this regard, Eq. (\ref{Gallagerbound}) is obtained to evaluate the error upper bound on Spinal codes.

Note that \cite{ref18} neglects the specific approach to evaluate the right-hand side (RHS) of Eq. (\ref{E0Q}) for Spinal codes, we also complemented a detailed approach to calculate it in Appendix A.

\subsubsection{The Bound Based on RCU Bound}
\begin{theorem} \label{lemma2}
	{\rm \textbf{(The upper bound on the error probability over the BSC \cite{ref18})} Consider Spinal codes with message length $n$ and segmentation parameter $k$ transmitted over a BSC with crossover probability $f$. Let $L$ be the number of passes the receiver has received. The average error probability under ML decoding can be upper bounded by}
	\begin{equation}\label{RCU}
	P_{e} \leq 1 - \prod_{a=1}^{n/k}(1-\epsilon_{B}(L_{a}, U_{a})),
	\end{equation}
	{\rm with}
	
	\begin{equation}\label{RCUbound}
	\epsilon_{B}(L_{a}, U_{a}) = \sum_{d=0}^{L_{a}}  \left\lbrace \tbinom{L_{a}}{d}  f^{d} \left( 1-f \right)^{L_{a}-d}    \times \min \left[   1, \left( U_{a}-1\right)\sum_{t=0}^{d}\tbinom{L_{a}}{t}2^{-L_{a}} \right]    \right\rbrace,\tag{\ref{RCU}a}
	\end{equation}
	{\rm where $L_a=L(n/k-a+1)$, $U_{a}=2^{k(n/k-a+1)}$. }
\end{theorem}
Theorem \ref{lemma2} utilizes the RCU bound to evaluate the performance of Spinal codes over the BSC. First proposed in \cite{ref20}, RCU bound is a significant general bounding technique that requires the property of pairwise independent property. As Spinal codes are also a type of pairwise independent random codes, the RCU bound is introduced to obtain Eq. (\ref{RCUbound}), thereby leading to the upper bound on the error probability of Spinal codes.
\subsection{A New Upper Bound over the AWGN Channel}
Different from the analysis in \cite{ref18}, which investigates the performance of Spinal codes by introducing general bounding techniques on random codes, we pay attention to Spinal code itself to obtain the new characterized upper bounds. Specifically, we systematically consider the coding structure, the decoding rule and the pairwise independent property of Spinal codes to conduct performance analysis for Spinal codes.  
\begin{theorem} \label{theorem5}
 {\rm \textbf{(The new upper bound on the error probability over the AWGN channel)} Consider Spinal codes with message length $n$, segmentation parameter $k$ and modulation parameter $c$ transmitted over an AWGN channel with noise variance $\sigma^{2}$. The average error probability under ML decoding can be upper bounded by}	
	\begin{equation}	\label{newbound} 
		P_{e} \le 1 - \prod_{a=1}^{n/k}(1-\min \left\{ {1, R_a} \right\}),
	\end{equation}
	{\rm with}
	\begin{equation} 
		R_{a}=\left( {{2^k} - 1} \right){2^{n - ak}}\frac{{{{ \pi  }^{{ L_a}/2}}}{ L_a}}{{\Gamma \left( {1 + \frac{{{ L_a}}}{2}} \right)}}\left( {\frac{{\int\limits_0^\mathcal{D} {{r^{2{ L_a} - 1}}} {e^{ - \frac{{{r^2}}}{{2{\sigma ^2}}}}}dr}}{{{{\left( {{2^c} - 1} \right)}^{{ L_a}}}\Gamma \left( {1 + \frac{{{ L_a}}}{2}} \right){{\left( {\sqrt {2{\sigma ^2}} } \right)}^{{ L_a}}}}} + \frac{{\int\limits_\mathcal{D}^\infty  {{r^{{ L_a} - 1}}} {e^{ - \frac{{{r^2}}}{{2{\sigma ^2}}}}}dr}}{{{{\left( {\sqrt {2\pi {\sigma ^2}} } \right)}^{{ L_a}}}}}} \right),\tag{\ref{newbound}a}
	\end{equation}
	\begin{equation} 
		\mathcal{D}=\frac{{\left( {{2^c} - 1} \right)\Gamma {{\left( {1 + { L_a}/2} \right)}^{1/{ L_a}}}}}{{\sqrt \pi  }},\tag{\ref{newbound}b}
	\end{equation}
	\begin{equation}
		L_a={\sum\limits_{i = a}^{n/k} {{\ell_i}} },\tag{\ref{newbound}c}
	\end{equation}
	{\rm where $\Gamma(\cdot)$ denotes the Gamma function, and $\ell_{i}$ is the number of transmitted symbols generated from the spine value $s_i$.}
\end{theorem}

Note that the right-hand side (RHS) of (\ref{newbound}a) can be immediately calculated by solving the integrals ${\int\limits_0^\mathcal{D} {{r^{2{ L_a} - 1}}} {e^{ - \frac{{{r^2}}}{{2{\sigma ^2}}}}}dr}$ and ${\int\limits_\mathcal{D}^\infty  {{r^{{ L_a} - 1}}} {e^{ - \frac{{{r^2}}}{{2{\sigma ^2}}}}}dr}$:

\begin{lemma}
	${\int\limits_0^\mathcal{D} {{r^{2{ L_a} - 1}}} {e^{ - \frac{{{r^2}}}{{2{\sigma ^2}}}}}dr}$.\\
	{\rm Let $\mathcal{I}_i\triangleq\prod\limits_{j = 1}^{i} {2\left( { L_a - j} \right)}$, the integral is calculated by: }
	\begin{equation}
		{\int\limits_0^\mathcal{D} {{r^{2{ L_a} - 1}}} {e^{ - \frac{{{r^2}}}{{2{\sigma ^2}}}}}dr}={\sigma ^{2 L_a}}\left( { - {e^{\frac{{ - { \mathcal{D}^2}}}{{2{\sigma ^2}}}}}\sum\limits_{i = 1}^ {L_a} {\frac{{{ \mathcal{D}^{2\left( { L_a - i} \right)}}}}{{{\sigma ^{2\left( { L_a - i} \right)}}}}} \mathcal{I}_{i-1} + \mathcal{I}_{L_a-1}  } \right).
	\end{equation}
\end{lemma}
\begin{proof}
	Please See Appendix B.
\end{proof}

\begin{lemma}${\int\limits_\mathcal{D}^\infty  {{r^{{ L_a} - 1}}} {e^{ - \frac{{{r^2}}}{{2{\sigma ^2}}}}}dr}$.\\
	{\rm Let $\mathcal{K}_i\triangleq\prod\limits_{j = 1}^{i} {\left( { L_a - 2j} \right)}$, the integral is calculated by: }
	\begin{equation}
		{\int\limits_\mathcal{D}^\infty  {{r^{{ L_a} - 1}}} {e^{ - \frac{{{r^2}}}{{2{\sigma ^2}}}}}dr}=\left\{
		\begin{array}{*{20}{c}}
		{{\sigma ^{L_a}}{e^{\frac{{ - {\mathcal{D}^2}}}{{2{\sigma ^2}}}}}\sum\limits_{i = 1}^{{L_a}/2} {\frac{{{\mathcal{D}^{\left( {{L_a} - 2i} \right)}}}}{{{\sigma ^{\left( {{L_a} - 2i} \right)}}}}} \mathcal{K}_{i-1} },&\text{if } L_a \text{ is even}\\
		{{\sigma ^{L_a}}\left( {\sqrt {2\pi } Q\left( {\mathcal{D}/\sigma } \right) \mathcal{K}_{\left(L_a-3\right)/2} + {e^{\frac{{ - {\mathcal{D}^2}}}{{2{\sigma ^2}}}}}\sum\limits_{i = 1}^{\left( {{L_a} - 1} \right)/2} {\frac{{{\mathcal{D}^{{L_a} - 2i}}}}{{{\sigma ^{{L_a} - 2i}}}}} \mathcal{K}_{i-1}  } \right)}, &\text{if } L_a \text{ is odd}
		\end{array}\right..
	\end{equation}
	{\rm where $Q\left(\cdot\right)$ is the $Q$ function defined as $Q\left(x\right)\triangleq\int_x^{ + \infty } {\frac{1}{{\sqrt {2\pi } }}} {e^{ - \frac{{{t^2}}}{2}}}dt$}.
\end{lemma}
\begin{proof}
	Please See Appendix C.
\end{proof}
\textbf{The proof of Theorem 3 is given as follows: }
\begin{proof}
	Suppose message $M = \left( {{m_1},{m_2} \ldots ,{m_{n/k}}} \right)\in\left\{0,1\right\}^n$ is encoded to Spinal code words $\operatorname{f}\left({x}_{i,j}\left(M\right)\right)$ to be transmitted through an AWGN channel. Let $y_{i,j}$ denote the corresponding received symbol at the receiver, and denote $n_{i,j}$ as the AWGN with variance $\sigma^2$. The received symbol $y_{i,j}$ is:
		\begin{equation}
		{y_{i,j}} = \operatorname{f}\left(x_{i,j}(M)\right){\rm{ + }}{n_{i,j}}.
		\end{equation} 
		
	The ML decoding process is to select the one with the lowest decoding cost from the candidate sequence space $\left\{0,1\right\}^n$. In this regard, we simply divide the candidate sequence space into two categories: $i$) the correct decoding sequence, i.e., $M$; $ii$) the wrong decoding sequences $M'=\left( {{m'_1},{m'_2} \ldots ,{m'_{n/k}}} \right)\in\mathcal{W}$, which are not unique and will form a set $\mathcal W $. The core idea of the ML decoding process is to distinguish theses two categories by their decoding costs. Thus, we begin with analyzing the decoding costs of them.
	
	The cost of the correct decoding sequence $M$ is defined as $D\left(M\right)$, calculated by
	\begin{equation}
	D(M)\triangleq\sum\limits_{i = 1}^{n/k} {\sum\limits_{j = 1}^{{\ell_i}} {({y_{i,j}}} }  - \operatorname{f}\left({{x_{i,j}}\left({M}\right)}\right){^2}\\
	= \sum\limits_{i = 1}^{n/k} {\sum\limits_{j = 1}^{{\ell_i}} {{n^2_{i,j}}} }.
	\end{equation}


	 Similarly, the cost of a wrong decoding sequence $M'\in\mathcal{W}$ is defined as $D\left(M'\right)$: 
	\begin{equation}
	D\left(M'\right)\triangleq\sum\limits_{i = 1}^{n/k} {\sum\limits_{j = 1}^{{\ell_i}} {({y_{i,j}}} }  - \operatorname{f}\left({{x_{i,j}}\left({M'}\right)}\right){^2}.
	\end{equation}
	
	Denote $E_{a}$ as the event that there exists an error in the $a^{\rm th}$ segment of Spinal codes and $ \bar E_{a}$ as the opposing event of $E_a$. The error probability of Spinal codes is:
	\begin{equation} \label{eq16}
	\begin{aligned}
	{P_e} &= {\mathbb{P}}\left( \bigcup_{a=1}^{n/k}{{E_a}} \right) = 1-{\mathbb{P}}\left( \bigcap_{a=1}^{n/k}{\bar {E_a}} \right)\\
	&= 1 - \prod\limits_{a = 1}^{n/k} \left[{1-{\mathbb{P}}\left({{{ E}_a}|{{\bar E}_1}, \cdots ,{{\bar E}_{a - 1}}} \right)}\right].
	\end{aligned}
	\end{equation}
	Define $\mathcal{W}_a\triangleq\left\{(m'_1, \dots, m'_{n/k}):m'_1 = m_1, \dots, m'_{a-1} = m_{a-1}, m'_a \ne m_a \right\} \subseteq \mathcal{W}$, which consists of wrong candidate sequences $M'$ with condition that event $\{{{{ E}_a}|{{\bar E}_1}, \cdots ,{{\bar E_{a - 1}}}}\}$ is true.
	
	Within these notations, $\mathbb{P}\left( {{{ E}_a}|{{\bar E}_1}, \cdots ,{{\bar E}_{a - 1}}} \right)$ is given as:
	\begin{equation}\label{conditionp}
	\begin{aligned}
		\mathbb{P}\left( {{{ E}_a}|{{\bar E}_1}, \cdots ,{{\bar E}_{a - 1}}}\right) &= \mathbb{P}\left( {\exists {M' \in \mathcal{W}_a}:D\left( M' \right) \le D\left( M \right)} \right). \\
	\end{aligned}
    \end{equation}
    Note that the RHS of (\ref{conditionp}) can be upper bounded by the union bound of probability, which yields the inequality:
    \begin{equation}\label{definition}
        \mathbb{P}\left( {\exists {M' \in \mathcal{W}_a}:D\left( M' \right) \le D\left( M \right)} \right)\le\sum\limits_{M' \in \mathcal{W}_a  } {\mathbb{P}\left( {D\left( {M'} \right) \le D\left( M \right)} \right)}.
    \end{equation}

Next, we turn our attention back to the the iterative process that $s_i=h\left(s_{i-1},m_i\right), s_0=\mathbf{0}^v$, which has been given in (\ref{spine}), to further analyze the RHS of (\ref{definition}). Denote the spine values of $M'$ and $M$ as $\mathbf{s}'=\left(s'_1,\dots, s'_{n/k}\right)$ and $\mathbf{s}=\left(s_1,\dots, s_{n/k}\right)$, respectively. Since $M'\in \mathcal{W}_a$, we obtain that $s'_i = s_i$ for any $i<a$. This leads to a further conclusion that  $\operatorname{f}\left(x_{i,j}\left(M'\right)\right)=\operatorname{f}\left(x_{i,j}\left(M\right)\right)$ for any $i<a$.  Denote $\mathbf{y}^{L_a}$ as the vector consisted of $y_{i,j}$ with $a \le i \le n/k, 1 \le j \le \ell_{i}$. Similarly defining $\mathbf{x}^{L_a}\left(M'\right)$ as the vector consisted of $\operatorname{f}\left(x_{i,j}\left(M'\right)\right)$ and $\mathbf{N}^{L_a}$ as the vector consisted of $n_{i,j}$ with $a \le i \le n/k, 1 \le j \le \ell_{i}$, we have
 \begin{equation}\label{integralform}
 \begin{aligned}
 &\sum\limits_{M' \in \mathcal{W}_a  } {\mathbb{P}\left( {D\left( {M'} \right) \le D\left( M \right)} \right)}\\
 &=\sum\limits_{M' \in \mathcal{W}_a  }\mathbb{P}\left( \sum\limits_{i = 1}^{n/k} {\sum\limits_{j = 1}^{{\ell_i}} {({y_{i,j}}}- \operatorname{f}\left(x_{i,j}\left(M'\right)\right){)^2} }   \le \sum\limits_{i = 1}^{n/k} {\sum\limits_{i = 1}^{{\ell_i}} n_{i,j}^2 }  \right)\\ 
 &=\sum\limits_{M' \in \mathcal{W}_a  }\mathbb{P}\left( \sum\limits_{i = a}^{n/k} {\sum\limits_{j = 1}^{{\ell_i}} {({y_{i,j}}}- \operatorname{f}\left(x_{i,j}\left(M'\right)\right){)^2} }   \le \sum\limits_{i = a}^{n/k} {\sum\limits_{j = 1}^{{\ell_i}} n_{i,j}^2 }  \right)\\
 &={\sum\limits_{M' \in \mathcal{W}_a  } \underbrace{\int_{\mathbb{R}}\dots\int_{\mathbb{R}}}_{L_a} \mathbb{P}\left( \left. {\sum\limits_{i = a}^{n/k} {\sum\limits_{j = 1}^{{\ell_i}} {({y_{i,j}}}- \operatorname{f}\left(x_{i,j}\left(M'\right)\right){)^2} }   \le \sum\limits_{i = a}^{n/k} {\sum\limits_{j = 1}^{{\ell_i}} n_{i,j}^2 }}\right|\mathbf{N}^{L_a} \right)} \prod_{i=a}^{n/k}\prod_{j=1}^{\ell_i}f\left(n_{i,j}\right)\prod_{i=a}^{n/k}\prod_{j=1}^{\ell_i}dn_{i,j}\\
 &={\sum\limits_{M' \in \mathcal{W}_a  } \underbrace{\int_{\mathbb{R}}\dots\int_{\mathbb{R}}}_{L_a} \mathbb{P}\left( \left. { \| \mathbf{y}^{L_a}-\mathbf{x}^{L_a}\left(M'\right) \| ^2  \le \sum\limits_{i = a}^{n/k}{\sum\limits_{j = 1}^{{\ell_i}} n_{i,j}^2 } }\right|\mathbf{N}^{L_a} \right)} \prod_{i=a}^{n/k}\prod_{j=1}^{\ell_i}f\left(n_{i,j}\right)\prod_{i=a}^{n/k}\prod_{j=1}^{\ell_i}dn_{i,j}.\\
 \end{aligned}
 \end{equation}
 where $n_{i,j}$ is the AWGN with distribution that 
 \begin{equation}\label{gaussian}
 f\left( {{n_{i,j}}} \right) = \frac{1}{{\sqrt {2\pi {\sigma ^2}} }} \cdot {e^{ - \frac{{n_{i,j}^2}}{{2{\sigma ^2}}}}}.
 \end{equation}
 
 Also, as $M'\in\mathcal{W}_a$ and because of the pairwise independent property of hash functions, the $a^{\rm th}$ spine value of the wrong decoding sequence $s'_a=h(m'_a,s'_{a-1})$ is independent with the $a^{\rm th}$ spine value of the correct decoding sequence $s_1=h(m_a,s_{a-1})$. Note that the spine values of Spinal codes are iteratively obtained with $s_i=h(m_i,s_{i-1})$. This leads to the pairwise independence of $s_i$ and $s'_i$ for all $a \le i \le n/k$. Thereby, the symbols generated by the RNG seeded with $s'_i$ are also independent with the symbols generated by the RNG seeded with $s_i$ for $a \le i \le n/k$, which yields that $\operatorname{f}\left(x_{i,j}\left(M'\right)\right)$ is independent with $\operatorname{f}\left(x_{i,j}\left(M\right)\right)$ and follows the uniform distribution $\mathcal{U}(0,2^c-1)$ for $a \le i \le n/k$. As such, the probability in (\ref{integralform}) is upper bounded by:
 
 \begin{equation}\label{eq10}
 \begin{aligned}
&\mathbb{P}\left( \left. { \| \mathbf{y}^{L_a}-\mathbf{x}^{L_a}\left(M'\right) \| ^2  \le \sum\limits_{i = a}^{n/k}{\sum\limits_{j = 1}^{{\ell_i}} n_{i,j}^2 } }\right|\mathbf{N}^{L_a} \right) \\
 &\le \frac{{\rm Vol}\left(\mathbb{B}^{L_a}\left( \sqrt{\sum\limits_{i = a}^{n/k}{\sum\limits_{j = 1}^{{\ell_i}} n_{i,j}^2 }}\right)\right)}{{\rm Vol}\left(\mathbb{C}^{L_a}\left(2^c-1\right)\right)}=\frac{{{{\left( {\pi \sum\limits_{i = a}^{n/k}\sum\limits_{j = 1}^{{\ell_i}} n_{i,j}^2} \right)}^{{L_a}/2}}}}{{{{\left( {{2^c} - 1} \right)}^{L_a}}\Gamma \left( {1 + \frac{L_a}{2}} \right)}}.
 \end{aligned}
 \end{equation}
 where 
 \begin{equation}
 \nonumber
 \begin{aligned}
 &\mathbb{B}^n\left(r\right)=\left\{\left(x_1,x_2,\cdots,x_n\right)\in \mathbb{R}^n \left| x_1^2+x_2^2+\cdots+x_n^2 \le r^2 \right. \right\},\\ 
 &\mathbb{C}^n\left(l\right)=\left\{\left(x_1,x_2,\cdots,x_n\right)\in \mathbb{R}^n \left| 0\le x_i \le l, \text{{\rm for} } i=1,\cdots,n\right. \right\},
 \end{aligned}
 \end{equation}
 ${\rm Vol}\left(\mathbb{B}^{n}\left(r\right)\right)=\frac{\pi^{n/2}}{\Gamma\left(1+\frac{n}{2}\right)}r^n$ represents the volume of the $n$-ball with radius $r$ \cite{vball}, and ${\rm Vol}\left(\mathbb{C}^{n}\left(l\right)\right)=l^n$ denotes the volume of the $n$-cube with side length $l$. 
 
Note that the volume of the $n$-ball in (\ref{eq10}) might be larger than the volume of the $n$-cube, while the probability is up to $1$, the function $\min \left\{{1, \bigcdot }\right\}$ can be applied to yield a further tightened inequality as follows:
  \begin{equation}\label{minfunction}
 \mathbb{P}\left( \left. { \| \mathbf{y}^{L_a}-\mathbf{x}^{L_a}\left(M'\right) \| ^2  \le \sum\limits_{i = a}^{n/k}{\sum\limits_{j = 1}^{{\ell_i}} n_{i,j}^2 } }\right|\mathbf{N}^{L_a} \right) \le \min\left\{1,\frac{{{{\left( {\pi \sum\limits_{i = a}^{n/k}\sum\limits_{j = 1}^{{\ell_i}} n_{i,j}^2} \right)}^{{L_a}/2}}}}{{{{\left( {{2^c} - 1} \right)}^{L_a}}\Gamma \left( {1 + \frac{L_a}{2}} \right)}}\right\}.
 \end{equation}
 Substituting (\ref{gaussian}) and (\ref{minfunction}) into (\ref{integralform}) results in the bound:
  \begin{equation}\label{bound2}
 \begin{aligned}
 &\sum\limits_{M' \in \mathcal{W}_a  } {\mathbb{P}\left( {D\left( {M'} \right) \le D\left( M \right)} \right)}\\
 &\le{\sum\limits_{M' \in \mathcal{W}_a  } \underbrace{\int\dots\int_{\mathbb{R}}}_{L_a} \min \left\{1,\frac{{{{\left( {\pi \sum\limits_{i = a}^{n/k}\sum\limits_{j = 1}^{{\ell_i}} n_{i,j}^2} \right)}^{{L_a}/2}}}}{{{{\left( {{2^c} - 1} \right)}^{L_a}}\Gamma \left( {1 + \frac{L_a}{2}} \right)}}\right\} 
	\frac{1}{{{{\left( {\sqrt {2\pi {\sigma ^2}} } \right)}^{L_a}}}} e ^{\frac{-\sum\limits_{i = a}^{n/k}\sum\limits_{j = 1}^{{\ell_i}} n_{i,j}^2}{2\sigma^2}} \prod_{i=a}^{n/k}\prod_{j=1}^{\ell_i}dn_{i,j}}.\\
 \end{aligned}
 \end{equation}

  Note that the RHS of (\ref{minfunction}) is a piecewise function. Let
 \begin{equation} \label{ball1}
 \nonumber
 \frac{{{{\left( {\pi \sum\limits_{i = a}^{n/k}\sum\limits_{j = 1}^{{\ell_i}} n_{i,j}^2} \right)}^{{L_a}/2}}}}{{{{\left( {{2^c} - 1} \right)}^{L_a}}\Gamma \left( {1 + \frac{L_a}{2}} \right)}}=1,
 \end{equation}
 we have
 \begin{equation}\label{boundd}
\sqrt{\sum\limits_{i = a}^{n/k}\sum\limits_{j = 1}^{{\ell_i}} n_{i,j}^2}=\frac{{\left( {{2^c} - 1} \right)\Gamma {{\left( {1 + { L_a}/2} \right)}^{1/{ L_a}}}}}{{\sqrt \pi  }}.
 \end{equation}
 Denote the RHS of (\ref{boundd}) as $\mathcal{D}$, and we obtain a piecewise form as follows:
 \begin{equation}\label{piecewise}
 \min \left\{1,\frac{{{{\left( {\pi \sum\limits_{i = a}^{n/k}\sum\limits_{j = 1}^{{\ell_i}} n_{i,j}^2} \right)}^{{L_a}/2}}}}{{{{\left( {{2^c} - 1} \right)}^{L_a}}\Gamma \left( {1 + \frac{L_a}{2}} \right)}}\right\}=\left\{
 \begin{array}{*{20}{c}}
 {1 }&\text{\rm if }\sqrt{\sum\limits_{i = a}^{n/k}\sum\limits_{j = 1}^{{\ell_i}} n_{i,j}^2} \ge \mathcal{D},\\
 \frac{{{{\left( {\pi \sum\limits_{i = a}^{n/k}\sum\limits_{j = 1}^{{\ell_i}} n_{i,j}^2} \right)}^{{L_a}/2}}}}{{{{\left( {{2^c} - 1} \right)}^{L_a}}\Gamma \left( {1 + \frac{L_a}{2}} \right)}} &\text{\rm if }\sqrt{\sum\limits_{i = a}^{n/k}\sum\limits_{j = 1}^{{\ell_i}} n_{i,j}^2} < \mathcal{D}.
 \end{array}\right.
 \end{equation}
 Plug (\ref{piecewise}) in (\ref{bound2}), we obtain the overall bound:
  \begin{equation}\label{boundoverall}
 \begin{aligned}
 &\sum\limits_{M' \in \mathcal{W}_a  } {\mathbb{P}\left( {D\left( {M'} \right) \le D\left( M \right)} \right)}\\
 &\le{\sum\limits_{M' \in \mathcal{W}_a  }	\frac{1}{{{{\left( {\sqrt {2\pi {\sigma ^2}} } \right)}^{L_a}}}} \idotsint\limits_{\sum\limits_{i = a}^{n/k}\sum\limits_{j = 1}^{{\ell_i}} n_{i,j}^2 \le \mathcal{D}^2}\frac{{{{\left( {\pi \sum\limits_{i = a}^{n/k}\sum\limits_{j = 1}^{{\ell_i}} n_{i,j}^2} \right)}^{{L_a}/2}}}}{{{{\left( {{2^c} - 1} \right)}^{L_a}}\Gamma \left( {1 + \frac{L_a}{2}} \right)}}
  e ^{\frac{-\sum\limits_{i = a}^{n/k}\sum\limits_{j = 1}^{{\ell_i}} n_{i,j}^2}{2\sigma^2}} \prod_{i=a}^{n/k}\prod_{j=1}^{\ell_i}dn_{i,j}}\\
 &+{\sum\limits_{M' \in \mathcal{W}_a  }	\frac{1}{{{{\left( {\sqrt {2\pi {\sigma ^2}} } \right)}^{L_a}}}} \idotsint\limits_{\sum\limits_{i = a}^{n/k}\sum\limits_{j = 1}^{{\ell_i}} n_{i,j}^2 \ge \mathcal{D}^2} e ^{\frac{-\sum\limits_{i = a}^{n/k}\sum\limits_{j = 1}^{{\ell_i}} n_{i,j}^2}{2\sigma^2}} \prod_{i=a}^{n/k}\prod_{j=1}^{\ell_i}dn_{i,j}}.\\
 \end{aligned}
 \end{equation}
We solve each of the terms in the above separately by introducing the hyperspherical coordinates:
\begin{equation}\label{sp1}
\begin{aligned}
&n_{a,1}=r\cos\phi_1,\\
&n_{a,2}=r\sin\phi_1\cos\phi_2,\\
&n_{a,3}=r\sin\phi_1\cos\phi_2\cos\phi_3,\\
&\vdots\\
&n_{n/k,\ell_{n/k}-1}=r\sin\phi_1\cdots\sin\phi_{L_a-2}\cos\phi_{L_a-1},\\
&n_{n/k,\ell_{n/k}}=r\sin\phi_1\cdots\sin\phi_{L_a-2}\sin\phi_{L_a-1}, 
\end{aligned}
\end{equation}
taking 
\begin{equation}\label{sp2}
\begin{aligned}
dV&=\prod_{i=a}^{n/k}\prod_{j=1}^{\ell_i}dn_{i,j}=\left|\det\left(\frac{\partial n_{i,j}}{\partial \left(r,\phi_\kappa\right)} \right)\right|_{1\le\kappa\le L_a-1} dr\prod_{i=1}^{L_a-1}d\phi_i, \\
&=r^{L_a-1}\prod_{i=1}^{L_a-2}\sin^{L_a-1-i}\left(\phi_i\right)dr\prod_{i=1}^{L_a-1}d\phi_i,\\
&0\le\phi_i\le \pi, {\rm \text{for }} i=1,2,\cdots,L_a-2,\\
&0\le\phi_{L_a-1}\le 2\pi.
\end{aligned}
\end{equation}
 Denote the first term in the RHS of (\ref{boundoverall}) as $\mathcal{J}_1$ and the second term as $\mathcal{J}_2$, applying (\ref{sp1}) and (\ref{sp2}) in \ref{boundoverall} yields the following equations:
\begin{equation}
\begin{aligned}
\nonumber
&\mathcal{J}_1={\sum\limits_{M' \in \mathcal{W}_a  }	\frac{1}{{{{\left( {\sqrt {2 {\sigma ^2}} } \right)}^{L_a}}}} \int_{0}^{2\pi}\int_{0}^{\pi}\dots\int_{0}^{\mathcal{D}}\frac{{{{r}^{{2L_a-1}}}}}{{{{\left( {{2^c} - 1} \right)}^{L_a}}\Gamma \left( {1 + \frac{L_a}{2}} \right)}}e ^{\frac{-r^2}{2\sigma^2}}dr \left( \prod_{i=1}^{L_a-2}\sin^{L_a-1-i}\left(\phi_i\right)d\phi_i \right)d\phi_{L_a-1}},\\
&\mathcal{J}_2={\sum\limits_{M' \in \mathcal{W}_a  }	\frac{1}{{{{\left( {\sqrt {2\pi {\sigma ^2}} } \right)}^{L_a}}}} \int_{0}^{2\pi}\int_{0}^{\pi}\dots\int_{\mathcal{D}}^{\infty} r^{L_a-1}e ^{\frac{-r^2}{2\sigma^2}} dr \left( \prod_{i=1}^{L_a-2}\sin^{L_a-1-i}\left(\phi_i\right)d\phi_i \right)d\phi_{L_a-1}}.
\end{aligned}
\end{equation}
Note that the integrals above can be decomposed as in
\begin{equation}\label{oveb}
\begin{aligned}
&\mathcal{J}_1={\sum\limits_{M' \in \mathcal{W}_a  }	\frac{\int_{0}^{\mathcal{D}}{r}^{{2L_a-1}}e ^{\frac{-r^2}{2\sigma^2}}dr} {{{{\left( {\sqrt {2 {\sigma ^2}} } \right)}^{L_a}{{{{\left( {{2^c} - 1} \right)}^{L_a}}\Gamma \left( {1 + \frac{L_a}{2}} \right)}}}}} \int_{0}^{2\pi}\int_{0}^{\pi}\dots\int_{0}^{\pi} \left( \prod_{i=1}^{L_a-2}\sin^{L_a-1-i}\left(\phi_i\right)d\phi_i \right)d\phi_{L_a-1}},\\
&\mathcal{J}_2={\sum\limits_{M' \in \mathcal{W}_a  }	\frac{\int_{\mathcal{D}}^{\infty} r^{L_a-1}e ^{\frac{-r^2}{2\sigma^2}} dr}{{{{\left( {\sqrt {2\pi {\sigma ^2}} } \right)}^{L_a}}}} \int_{0}^{2\pi}\int_{0}^{\pi}\dots\int_{0}^{\pi} \left( \prod_{i=1}^{L_a-2}\sin^{L_a-1-i}\left(\phi_i\right)d\phi_i \right)d\phi_{L_a-1}},
\end{aligned}
\end{equation}
wherein the multiple integrals on the RHS is given by \cite{vball}: 
\begin{equation}\label{integral}
\int_{0}^{2\pi}\int_{0}^{\pi}\dots\int_{0}^{\pi} \left( \prod_{i=1}^{L_a-2}\sin^{L_a-1-i}\left(\phi_i\right)d\phi_i \right)d\phi_{L_a-1}=\frac{{{{ \pi  }^{{ L_a}/2}}}{ L_a}}{{\Gamma \left( {1 + \frac{{{ L_a}}}{2}} \right)}}.
\end{equation}
Applying (\ref{integral}) in (\ref{oveb}) and substituting (\ref{oveb}) into (\ref{boundoverall}) yields that

\begin{equation} \label{bound4}
\begin{aligned}
&\sum\limits_{M' \in \mathcal{W}_a  } {\mathbb{P}\left( {D\left( {M'} \right) \le D\left( M \right)} \right)} \\
&=\left| \mathcal{W}_a \right|\frac{{{{ \pi  }^{{ L_a}/2}}}{ L_a}}{{\Gamma \left( {1 + \frac{{{ L_a}}}{2}} \right)}}\left( {\frac{{\int\limits_0^\mathcal{D} {{r^{2{ L_a} - 1}}} {e^{ - \frac{{{r^2}}}{{2{\sigma ^2}}}}}dr}}{{{{\left( {{2^c} - 1} \right)}^{{ L_a}}}\Gamma \left( {1 + \frac{{{ L_a}}}{2}} \right){{\left( {\sqrt {2{\sigma ^2}} } \right)}^{{ L_a}}}}} + \frac{{\int\limits_\mathcal{D}^\infty  {{r^{{ L_a} - 1}}} {e^{ - \frac{{{r^2}}}{{2{\sigma ^2}}}}}dr}}{{{{\left( {\sqrt {2\pi {\sigma ^2}} } \right)}^{{ L_a}}}}}} \right),
\end{aligned}
\end{equation}
where $\left| \mathcal{W}_a \right|$ denotes the size of set $\mathcal{W}_a$ with $\left| \mathcal{W}_a \right| = \left( {{2^k} - 1} \right){2^{n - ak}}$. Now, by applying (\ref{bound4}) in (\ref{conditionp}) and (\ref{definition}) and denoting the RHS of (\ref{bound4}) as $R_a$, we have the inequality that
\begin{equation}\label{conditibound}
\mathbb{P}\left( {{{ E}_a}|{{\bar E}_1}, \cdots ,{{\bar E}_{a - 1}}}\right) \le \min \left\{1, R_a\right\}, 
\end{equation}
	
Eventually, substituting (\ref{conditibound}) into (\ref{eq16}) results in the derived bound.
\end{proof}

\subsection{A New Upper Bound over the BSC}

\begin{theorem} \label{theorem6}
	{\rm \textbf{(The new upper bound on the error probability over the BSC)} Consider Spinal codes with message length $n$ and segmentation parameter $k$ transmitted over a BSC with crossover probability $p$. Let $l_{j}$ be the number of symbols generated from the $j^{th}$ spine value of the $n$-bit message sequence. The FER of Spinal codes under ML decoding can be upper bounded by}
	\begin{equation} \label{BSCb}
	P_{e} \leq 1 - \prod_{a=1}^{n/k}(1-\epsilon_{a}) ,
	\end{equation}	
	{\rm with}
	\begin{equation} 
	{\epsilon _a} = \sum\limits_{d = 0}^{{L_a}} {\left( {\left( {\begin{array}{*{20}{c}}
				{{L_a}}\\
				d
				\end{array}} \right){p^d}{{\left( {1 - p} \right)}^{{L_a} - d}} \cdot \min \left\{{1,{R_{a,d}}}\right\}  } \right)}.\tag{\ref{BSCb}a}
	\end{equation}
	{\rm Note that}
	\begin{equation} 
	{R_{a,d}} = \left( {{2^k} - 1} \right){2^{n - ak}}\sum\limits_{t = 0}^d {\left( {\begin{array}{*{20}{c}}
			{{L_a}}\\
			t
			\end{array}} \right)} {2^{ - {L_a}}},\tag{\ref{BSCb}b}
	\end{equation}
	{\rm where}
	\begin{equation} 
	{L_a} = \sum\limits_{i = a}^{n/k} {{\ell_i}}.\tag{\ref{BSCb}c}
	\end{equation}		
\end{theorem}
	\begin{proof}
		Please refer to Appendix D.
	\end{proof}

\section{Transmission Scheme Design}
\label{section4}
This section aims to optimize the transmission scheme of Spinal codes. We begin with formulating and solving a rate maximization problem under ML decoding, wherein the derived upper bounds in Section \ref{section3} are adopted as a constraint. The results show that the maximum rate can be achieved by the incremental tail transmission scheme under ML decoding. Then, heuristically, we focus on the more practical near-ML bubble decoding algorithm. Under the bubble decoding algorithm, we design an improved transmission scheme named TITT scheme. The simulation results given in Section \ref{section6} show that the proposed TITT scheme enables further rate improvement for Spinal codes.
\subsection{Problem Formulation and Solution}
We establish a rate maximization model under a sufficiently small error probability constraint. Two related fundamental parameters are emphasize here:
\begin{itemize}
	\item \textit{Code rate}: Consider Spinal codes with message length $n$ and segmentation parameter $k$ transmitted over a channel. Let $\ell_{j}$ be the number of symbols generated from the $j^{th}$ spine value of the $n$-bit message sequence. The code rate can be calculated as
	\begin{equation} 
	\nonumber
		R=\frac{n}{\sum_{j=1}^{n/k}{\ell_{j}}}.
	\end{equation}
	
	\item \textit{Error Probability}: The error probability can be upper bounded by the results given in Theorem 3 and Theorem 4 over the AWGN channel and the BSC respectively. 
\end{itemize}

{\rm The rate maximization problem can be expressed as follows:}

	{\rm 1) Objective function: To maximize the code rate $R$.} 
	
	{\rm 2) Constraint: To ensure that the transmitted symbols can meet a prescribed requirement on successful decoding level that $P_{e} \le \delta$, we hold a stronger setting that: }
	\begin{equation}\nonumber
	P_{e}^{upper} = 1 - \prod_{a=1}^{n/k}(1-\epsilon_{a}) \le \delta,
	\end{equation}
	where $P_{e}^{upper}$ denotes the upper bound on the error probability for ML decoded Spinal codes, the value of $\delta$ can be preset according to specific application requirements.
	
	{\rm 3) Decision variables: The number of symbols generated from each spine value,} $\mathbf{L}= [\ell_{1}, \ell_2, \dots, \ell_{n/k}]$.

To sum up, the overall optimization problem can be formulated as follows.

\begin{problem}\label{p1}
{\rm For Spinal codes transmitted over the AWGN channel:} 
\begin{equation}\nonumber
\begin{aligned}
\max_{\mathbf{L}} \quad & R  \\
\mbox{\rm s.t.} \quad
		&c1:{{\ell_i} \in {\mathbb{Z}^ + }}, i=1,2,\dots,n/k,\\
		&c2:{P_e^{upper} = 1 - \prod\limits_{a = 1}^{n/k} {\left( {1{\rm{ - }}{\epsilon _a}} \right) \le {\delta }} },\\
		&c3:{{\epsilon _a} = \min \left\{1,R_a\right\}}.
\end{aligned}
\end{equation}
{\rm where $R_{a}$ is given in Theorem 3 and ${\mathbb{Z}^ +}$ stands for the positive integer set.}
\end{problem}

\begin{problem}\label{p2}
{\rm For Spinal codes transmitted over the BSC:}
\begin{equation}\nonumber
\begin{aligned}
\max_{\mathbf{L}}\quad & R  \\
\mbox{\rm s.t.} \quad  
	&c1:{\ell_i} \in {\mathbb{Z}^ +}, i=1,2,\dots,n/k,\\
	&c2:{P_e^{upper} = 1 - \prod\limits_{a = 1}^{n/k} {\left( {1{\rm{ - }}{\epsilon _a}} \right) \le {\delta }} },\\
	&c3:{{\epsilon _a} = \sum\limits_{d = 0}^{{L_a}} {\left( {\left( {\begin{array}{*{20}{c}}
						{{L_a}}\\
						d
						\end{array}} \right){f^d}{{\left( {1 - f} \right)}^{{L_a} - d}} \cdot \min \left\{{1,{R_{a,d}}}\right\} } \right)} }.
\end{aligned}
\end{equation}
{\rm where $R_{a,d}$ and $L_{a}$ are provided in Theorem 4.}	
\end{problem}
The above problems can be solved using nonlinear integer programming techniques. The Branch-Bound algorithm is a general method to solve this type of problem. However, the iterative mode of the Branch-Bound method suffers from high complexity and is likely to converge to a locally optimal solution. Motivated by the above, we design an algorithm customized for the problems above, which can solve the problem globally.

The general idea is to consider the dual problems rather than solve them directly. The dual problem of Problem \ref{p1} (or Problem \ref{p2}) is as follows.
\begin{problem}\label{p3}
	{\rm The dual problem of \textbf{Problem \ref{p1}} (or \textbf{Problem \ref{p2}}) :}
\begin{equation}\nonumber
\begin{aligned}
\min_{\mathbf{L}} \quad & P_e^{upper} \\
\mbox{\rm s.t.} \quad 
		&c1:{\sum_{i=1}^{n/k}\ell_i {{ = N}}},\\
		&c2:{{\ell_i} \in {\mathbb{Z}^ +}}, i=1,2,\dots,n/k.
\end{aligned}
\end{equation}
\end{problem}

In essence, the dual problem above is to determine the decision variables $\mathbf{L}= [\ell_{1}, \ell_2, \dots, \ell_{n/k}]$ under the constraint of a fixed code rate $n/{N}$, so as to minimize the error probability of the transmission scheme. Though the difference between the dual problem and the original problem is just a simple exchange of the constraint and the objective function, the constraint corresponding to the decision variables in the dual problem is a linear constraint, which is much simpler than the original problem.

We devise an on-the-fly algorithm to solve Problem \ref{p1} or Problem \ref{p2} by leveraging their dual problems (see Algorithm \ref{Algorithm 1}). By `on-the-fly', the minimum error probability is found by repeatedly solving Problem \ref{p3} with an increasing number of transmitted symbols. The on-the-fly process is terminated to output $\mathbf{L}$ once the upper bound on the error probability meets the condition that $P_e^{upper} \le \delta$. As the upper bound on the error probability of Spinal codes is monotonically decreasing with $N$, the output $\bf L$ also maps to a minimum number of cumulative transmission symbols $N$ such that $P_e^{upper} \le \delta$. Note that the minimum number of transmission symbols corresponds to the maximum code rate, the problem of rate maximization under the constraint of a prescribed small error probability $\delta$ is thereby solved out. 
\begin{algorithm}
	\label{Algorithm 1}
	\caption{The optimal algorithm for solving Problem \ref{p1} (or Problem \ref{p2})}
	\LinesNumbered
	\KwIn{Initial number of transmitted passes $r$; Preset reliability target $\delta$;}
	\KwOut{The number of symbols generated from each spine value $\mathbf{L}= [\ell_{1}, \ell_2, \dots, \ell_{n/k}]$;}
	\While{$P_e^{upper} \ge {\delta}$ }
	{
		Update the number of transmitted symbols: $N \leftarrow N + 1$\;
		Solve the Problem \ref{p3} to find out the minimum FER $P_{e,N}^{upper}$\;
		Update the minimum FER upper bound $P_e^{upper} \leftarrow P_{e,N}^{upper}$\;
	}
	\Return $\bf L$
\end{algorithm}

 We initially used Algorithm \ref{Algorithm 1} to obtain the optimal solution to Problem 1. With these results in hand, we try to design a greedy baseline algorithm to solve this problem, which significantly reduces the calculations caused by repeatedly solving problem \ref{p3}. Our experiments showed no difference in solution results between these two algorithms. Specifically, the greedy baseline algorithm always determines the current most error-prone symbol, that is, the symbol that can minimize the current FER to transmit. The details of the greedy baseline algorithm are given in Algorithm \ref{Algorithm 2}.

\begin{algorithm}
	\label{Algorithm 2}
	\caption{The greedy baseline algorithm for solving Problem \ref{p1} (or Problem \ref{p2})}
	\LinesNumbered
	\KwIn{Initial number of transmitted passes $r$; Preset reliability target $\delta$;}
	\KwOut{The number of symbols generated from each spine value $\mathbf{L}= [\ell_{1}, \ell_2, \dots, \ell_{n/k}]$;}
		 Initialization: $\mathbf{L} = [r,r, \ldots ,r,r]$, $N = rn/k$\;
		 Calculate $P_e^{upper}$ by applying Theorem 3 (for problem 1) or Theorem 4 (for problem 2)\;
		\While{$P_e^{upper} \ge {\delta}$ }
		{
			Update the number of transmitted symbols: $N \leftarrow N + 1$\;
			\For{$i \leftarrow 1$ to $n/k$}
			{
				 Update the decision variable: $\ell_{i} \leftarrow \ell_{i} + 1$\;
				 Calculate and store the corresponding FER $P_{e,i}^{upper}$\;
			     Restore the decision variable: $\ell_{i} \leftarrow \ell_{i} - 1$\;
		    }
				 Search for the minimum $P_{e,i}^{upper}$ and the corresponding index $d$\;
				 $\ell_{d} \leftarrow \ell_{d} + 1$, $P_e^{upper} \leftarrow P_{e,d}^{upper}$\;
        }
		\Return $\bf L$
\end{algorithm}

\begin{remark} \label{rm4}
	{\rm The value of $\delta$ can be preset based on specific application requirements. For example, typical ultra-reliable and low latency communication (URLLC) systems commonly require the error probability lower than $10^{-5}$. In our simulations, we set $\delta=10^{-5}$.}	
\end{remark}


\subsection{Improved Transmission Scheme under ML Decoding}
 Consider Spinal codes with $n=32$ and $k=4$, initially sending several passes is generally used for the decoder to construct an integral decoding tree. In this paper, we set $r=2$ and $r=3$ as case studies. By using both Algorithm \ref{Algorithm 1} and Algorithm \ref{Algorithm 2}, we obtain exactly the same solutions as shown in Table \ref{table1}.

\begin{table}[H] 
	\centering
	\caption{}\label{table1}
	Solutions to the Rate Maximization Problem\\
	(a) Transmission Scheme over the BSC\\
	\begin{tabular}{ccccccccc}
		Crossover probability &$r=2$ & $r=3$ \\ \hline
		0.05 & { ${\bf L} = [2,2,2,2,2,2,2,49]$}& {${\bf L} = [3,3,3,3,3,3,3,42]$}  \\
		{0.01} & { ${\bf L} = [2,2,2,2,2,2,2,40]$} &{${\bf L} = [3,3,3,3,3,3,3,33]$}  \\
		{0.005} & {${\bf L} = [2,2,2,2,2,2,2,34]$} &{${\bf L} = [3,3,3,3,3,3,3,28]$}  \\
		{0.001} & { ${\bf L} = [2,2,2,2,2,2,2,27]$} &{${\bf L} = [3,3,3,3,3,3,3,21]$} \\	\hline \\
	\end{tabular}\\					
	{(b) Transmission Scheme over the AWGN channel}\\
	\begin{tabular}{ccccccccc}
		{SNR (dB)} &{$r=2$} & {$r=3$} \\ \hline
		{7} & {${\bf L} = [2,2,2,2,2,2,2,34]$} &${\bf L} = [3,3,3,3,3,3,3,29]$  \\
		{8} & {${\bf L} = [2,2,2,2,2,2,2,25]$} &${\bf L} = [3,3,3,3,3,3,3,21]$  \\
		{9} & {${\bf L} = [2,2,2,2,2,2,2,19]$} &${\bf L} = [3,3,3,3,3,3,3,17]$  \\
		{10} & {${\bf L} = [2,2,2,2,2,2,2,16]$} &${\bf L} = [3,3,3,3,3,3,3,14]$  \\ \hline
	\end{tabular}
\end{table}

It can be observed from Table \ref{table1} that with ML decoding algorithm, the transmitter tends to continuously transmit the tail symbols, i.e., the symbols generated from the ${(n/k)}^{\rm th}$ spine value. This `incremental-tail-transmission'-pattern result can be explained and generalized universally by the serial coding structure of Spinal codes. Due to the serial coding structure with $s_i=h(s_{i-1},m_i)$, the symbols generated from the $i^{\rm th}$ spine value $s_i$ is related to the block message $m_i$ and the previous spine value $s_{i-1}$. Since $s_i$ is related to $s_{i-1}$ and $s_{i-1}$ is related to the block message $m_{i-1}$, we can conclude that the symbols generated from the $i^{\rm th}$ spine value $s_i$ is related to all $m_a$ with $a \le i$, which also means that for Spinal codes, tail symbols generated from $s_{n/k}$ are more relevant to the whole message sequence, i.e., $m_a$ with $a \le n/k$, and hence carries more information. 

Through empirical observations and qualitative analysis, we can conclude that the transmitting tail symbols can further improve the code rate of ML decoded Spinal codes.

\begin{remark} \label{rm5}
	{\rm The ML decoding algorithm is difficult to be applied in a practical communication system due to its high complexity. Though not applicable, it provides an ideal benchmark performance of Spinal codes. Furthermore, the related analysis and transmission scheme design can serve as an inspirational basis for the transmission scheme design with some near-ML decoding algorithms. }
\end{remark}

\subsection{Improved Transmission Scheme under Bubble Decoding}

In this subsection, inspired by the incremental tail transmission scheme for ML decoding, we propose an improved transmission scheme named thresholded incremental tail transmission scheme (TITT scheme). The TITT scheme combines uniform puncturing with incremental tail transmission scheme for the practical Spinal codes using bubble decoding algorithm.

As introduced in Section \ref{section2c}, bubble decoding is a kind of  near-ML decoding algorithm. It prunes the branch at each layer and retains only $B$ nodes with the lowest cost. It is obvious that the decoding will succeed only if the decoder prunes the branch correctly at every layer of the decoding tree.

Recall that the cost can be calculated in (\ref{layer}). Since the calculation of the cost at each layer is only related to the preceding received symbols, i.e., $y_{i,j}$ with $i \le \lambda$, the inadequate transmission of the preceding symbols may result in an incorrect pruning, and thus directly leads to a decoding error. Therefore, for bubble decoding algorithm, merely transmitting the tail symbols without considering the sufficiency of transmission of preceding symbols is not rational. In addition, in the pruning process of the bubble decoding algorithm, the previous layer will retain $B$ candidate sequences, while the last layer will retain only one candidate sequence, which means that the pruning of the bubble algorithm in the last layer is more prone to errors.

\begin{algorithm}
	\caption{The TITT scheme for bubble decoding}
	\label{Algorithm 3}
	\LinesNumbered
	\KwIn{The transmission order vector ${\bf g} = [g_{1},g_{2},\ldots,
		g_{n/k}]$; The switch threshold $T_{\rm r}$;}
		 Transmit a complete pass for the decoder to build an integral decoding tree: $j \leftarrow 1$, $N \leftarrow n/k$\;
		\While {not receiving an ACK}
		{
			 $ j\leftarrow j+1$\;			
			 \If{$N \le {T_{r}}$}
			 {
			 	$ i\leftarrow 1$\;
	         	\While {$i \le  n/k$ and $N < T_r$ }
	         	{
	         		Transmit coded symbol $\operatorname{f}(x_{g_i,j}(M))$\;
			 		$N \leftarrow N+1$\;
			 		$i \leftarrow i+1$\;
			 	}
		 	}
			 	\Else {Transmit tail symbol $\operatorname{f}(x_{n/k,j}(M))$\;}	
		}
\end{algorithm}

Considering the analysis above, we propose an improved transmission scheme for bubble decoding in Algorithm \ref{Algorithm 3}. It combines uniform puncturing with incremental tail transmission, forming a two-stage scheme. The first stage adopts the uniform-puncturing-based transmission, aiming to refine the code rate granularity and more importantly, to ensure the pruning at the preceding levels of the decoding tree is correct. After sufficient symbols having been sent, the transmitter switches to the second stage, which adopts the incremental tail transmission scheme, aiming to improve the probability that the correct candidate sequence is selected within the $B$-node list at the last layer. Let $T_{r}$ denote the switch-over threshold of the number of transmitted symbols. The detailed procedures of the improved transmission scheme for bubble decoding are described in Algorithm \ref{Algorithm 3}.

The value of the switch-over threshold $T_{r}$ in Algorithm \ref{Algorithm 3} can be set according to the following inferences: with increasing number of transmitted symbols $N$, the instantaneous code rate of the rateless code $R = n/N$ decreases; when the instantaneous code rate approaches the channel capacity $C$, it is highly probable that the previous pruning is correct. Therefore, we can set $T_{r}\approx n/C $. In this work, we provide the following heuristic settings, by which we have obtained satisfactory simulation performance, as will be shown in Section \ref{section6}.
\begin{itemize}
	\item For the AWGN channel,
	\begin{equation}\nonumber
	{T_{r}}=\left\lfloor {\frac{n}{{{C_{AWGN}}}} - \frac{n}{k}} \right\rfloor.
	\end{equation}
	\item For the BSC,
	\begin{equation}\nonumber
	{T_{r}}=\left\lfloor {\frac{n}{{{C_{BSC}}}}} \right\rfloor.
	\end{equation}
\end{itemize}
%

\section{Bubble Decoding with Memory (BD-M)}
\label{section5}
In essence, both uniform puncturing and the proposed TITT scheme improve the rate performance by refining the code rate granularity of Spinal codes. However, one trade-off is that the refinement of code rate granularity increases the number of decoding attempts at the receiver, resulting in extensive calculations. To address this issue, we investigate the dynamic changes of the decoding tree between adjacent decoding rounds of rateless Spinal codes and find that only parts of the decoding tree will change during the dynamic transmission-decoding process (see Theorem \ref{t5}). Therefore, the conventional full-tree updating process can be improved as a partial-tree updating one.  

\begin{theorem}\label{t5}
	{\rm If a symbol $y_{i,j}$ is newly received at the decoder, then only the decoding tree with layer $\lambda \ge i$ will change in the decoding process.}
\end{theorem}
\begin{proof}
Please refer to Appendix E.	
\end{proof}

Based on Theorem \ref{t5}, we know that if the receiver end introduces an extra memory to dynamically store and partially update the decoding tree of Spinal codes, the calculations required for the unchanged parts of the decoding tree can be eliminated, thereby reducing the decoding complexity. Inspired by this idea, we take the most commonly used bubble decoding algorithm as an example, and propose the bubble decoding with memory (BD-M) algorithm here. An interesting issue is that the proposed BD-M algorithm matches the proposed TITT scheme well. The combination of the TITT scheme at the transmitter and the BD-M algorithm at the receiver constitutes an improved transmission-decoding system, which not only exhibits good rate performance but also improves the decoding efficiency of Spinal codes.

\subsection{Decoding Process of the BD-M Algorithm}
Conventional bubble decoding is a full-tree-updating based algorithm, i.e., it needs to repeatedly reconstruct the integral decoding tree once receiving any new sub-passes or symbols. However, by introducing a decoding tree memory to the receiver, the decoder can store the constructed decoding tree and updates only a part of it. Based on this finding, we design the BD-M algorithm, which is a partial-tree-updating based algorithm. 

The primary decoding process of the BD-M algorithm can be summarized in four steps:
\begin{enumerate}
	\item Construct a pruned decoding tree according to the received integral pass. The pruning strategy is referred to the original bubble decoding algorithm \cite{Spinalcodes}.
	\item Store the pruned decoding tree.
	\item If the decoder receives symbol $y_{i,j}$, it will retain the top $i-1$ layers of the decoding tree and update the information after it according to the pruning strategy of bubble decoding algorithm.
	\item Repeat steps 2) and 3) until the decoding process is finished.
\end{enumerate}

It can be observed that the proposed BD-M algorithm does not change the bubble decoding-based pruning decision at each layer since the cost of each node calculated by (\ref{layer}) remains unchanged. The core idea of the BD-M algorithm is to restore the unchanged tree information and reduce the cost recalculation of the unchanged part.

\begin{remark} \label{rm6}
	{\rm The BD-M algorithm can be applied to both uniform puncturing and the proposed TITT scheme to reduce decoding complexity, matches better with the proposed TITT scheme. As described in Algorithm 3, the TITT scheme tends to transmit more tail symbols. Correspondingly, the BD-M algorithm at the decoder only needs to update the last layer of the decoding tree when receiving tail symbols. By this means, this algorithm dramatically eliminates the need for repeatedly calculating the decoding cost, as the only thing for the decoder is to update the last layer of the decoding tree. }
\end{remark}

\subsection{Qualitative Complexity Comparison }

During the decoding process, the computation involved at each layer of the decoding tree mainly includes the cost calculating and the bubble sorting for all the nodes at the layer. Let $x$ denote the number of nodes at a particular layer and $g(x)$ represent the associated computation amount.

First, we analyze the number of nodes at each decoding tree layer. According to the decoding process, the branch pruning starts if ${2^{ik}} \ge B$, where $i$ denotes the layer index of the decoding tree. Thus, we have
\begin{equation}\nonumber
\begin{aligned}
{x_i} = \left\{ {\begin{array}{*{20}{c}}
	{{2^{ik}}}&{1 \le i \le \left\lceil {\frac{{{{\log }_2}B}}{k}} \right\rceil }\\
	{B{2^k}}&{\left\lceil {\frac{{{{\log }_2}B}}{k}} \right\rceil  < i \le \frac{n}{k}}
	\end{array}} \right.,
\end{aligned}
\end{equation}
where $x_{i}$ denotes the number of nodes at the $i^{\rm th}$ layer.

Let $o_{i}$ denote the amount of calculations required to update the decoding tree from the $i^{\rm th}$ layer, we have:
\begin{equation}\nonumber
{o_i} = \sum\limits_{j = i}^{n/k} {g\left( {{x_j}} \right)}.
\end{equation}
It is obvious that ${o_{\rm{1}}} > {o_2} \ldots  > {o_{n/k}}$. $o_{1}$ denotes the computation amount required to reconstruct the full decoding tree.

Next, we analyze the total computation amount required by different transmission-decoding pairs and compare them.

\textit{1) Uniform Puncturing with Bubble Decoding vs. Uniform Puncturing with BD-M:}

For uniform puncturing (UP) with the original bubble decoding (BD) algorithm, the decoder needs to reconstruct the full decoding tree when receiving any sub-passes or symbols, so the total computation amount can be expressed as
\begin{equation}\label{eq17}
{O_{{\rm{UP}} \to {\rm{BD}}}} = \sum\limits_{i = 1}^{n/k} {{\ell_i}} {o_1},
\end{equation}
where $\ell_{i}$ is the number of transmitted symbols generated from the $i^{\rm th}$ spine value required to successfully recover the message when using the uniform-puncturing-based transmission scheme.

For uniform puncturing with the BD-M algorithm, the decoder will not reconstruct the full decoding tree but retain the existing decoding tree in memory and update only a part. The corresponding total computation amount can be expressed as
\begin{equation}\label{eq18}
{O_{{\rm{UP}} \to {\rm{BD-M}}}} =\sum\limits_{i = 1}^{n/k} {{\ell_i}} {o_i}.
\end{equation}

Since ${o_{\rm{1}}} > {o_2} \ldots  > {o_{n/k}}$, we can subtract (\ref{eq17}) from (\ref{eq18}) and obtain that ${O_{{\rm{UP}} \to {\rm{BD}}}}>{O_{{\rm{UP}} \to {\rm{BD-M}}}}$, showing that the proposed BD-M algorithm requires less computation than the original bubble decoding.

\textit{2) Uniform Puncturing with BD-M vs. TITT with BD-M:}
\begin{figure}[htbp]
	\centering
	{\includegraphics[angle=0,width=0.8\textwidth]{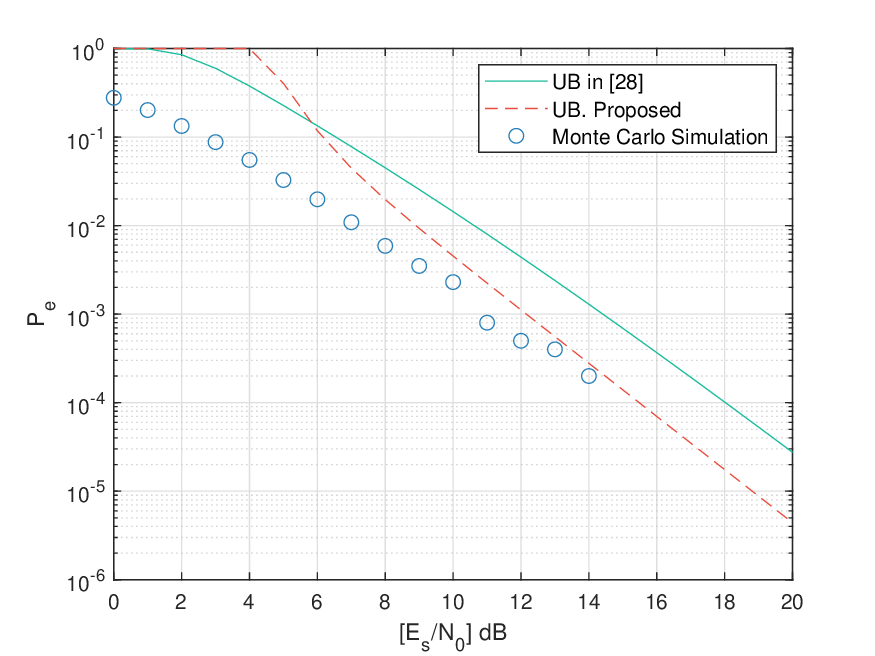}}	
	\caption{Upper bounds on the error probability of Spinal codes with $n=8$, $v=32$, $Pass=6$, $c=8$ and $k=2$ over the AWGN channel.
	}\label{Fignewupperbound}
\end{figure}
Let $\ell_{i}'$ denote the number of transmitted symbols generated from the $i^{\rm th}$ spine value required to successfully recover the message when using the proposed improved transmission scheme. The total computation amount of the TITT with BD-M can be expressed as
\begin{equation}\label{eq19}
{O_{{\rm{TITT}} \to {\rm{\ BD-M}}}} = \sum\limits_{i = 1}^{n/k} {{\ell_i}'{o_i}},
\end{equation}
Compare (\ref{eq18}) with (\ref{eq19}), we can prove that (see Appendix F)
\begin{equation}\label{eq20}
\sum\limits_{i = 1}^{n/k} {{\ell_i}{o_i}}  \ge \sum\limits_{i = 1}^{n/k} {{\ell_i}'{o_i}},
\end{equation}
which demonstrates the superiority of the proposed transmission scheme. As a result, the transmitter's improved transmission scheme and the matching BD-M algorithm at the decoder consist of an optimized transmission system for Spinal codes. 
\section{Simulation Results}
\label{section6}
In this section, we conduct Monte Carlo simulations to verify the upper bounds derived in Section \ref{section3} and demonstrate the gain of rate performance obtained by leveraging the TITT scheme. Also, we would show that the BD-M algorithm proposed in Section \ref{section5} matches well with the proposed TITT scheme.
\begin{figure}[htbp]
	\centering
	{\includegraphics[angle=0,width=0.8\textwidth]{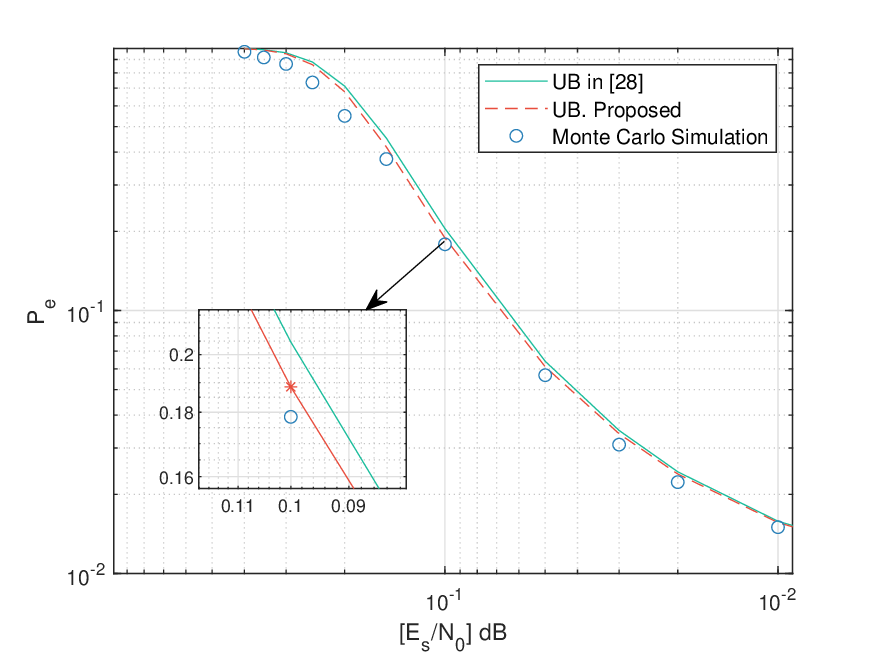}}
	\caption{Upper bounds on the error probability of Spinal codes with $n=8$, $v=32$, $Pass=8$, $c=1$ and $k=2$ over the BSC.}\label{FER comparision}
\end{figure}
\subsection{Bounding Performance}
 Fig. \ref{Fignewupperbound} is conducted over the AWGN channel model, which demonstrates the comparison among the upper bound proposed in \cite{ref18}, the upper bound given in (\ref{newbound}), and the Monte Carlo simulation results. Considering that the decoding complexity of ML decoding increases exponentially with $n$, we select $n$ as short as $n=8$ for the simulation setup. We set the the Monte Carlo Simulation sample size as $10^4$ and obtain the simulated error probability as in Fig. \ref{Fignewupperbound}. From Fig. \ref{Fignewupperbound}, we can observe that compared to the available bounds in \cite{ref18}, the proposed upper bound draws its strength from the tightness at high SNRs. However, at low SNRs, the proposed bound performs poor tightness. This phenomenon are mainly due to (\ref{eq10}), where the bound on the probability is given as the volume of a $L_a$-ball divided by the volume of a $L_a$-cube. Note that the radius of the hypersphere is exactly the $\ell_2$-norm of the noise vector, while the volume of the cube is a fixed value. In such a case, if the SNR is low, i.e., the noise variance is high, the RHS of (\ref{eq10}) may exceed $1$ and the bound will degenerate into an invalid one, with the LHS of (\ref{eq10}) upper bounded by $1$. However, in the high-SNR regime, the RHS of (\ref{eq10}) remains far away from $1$, thereby exhibiting the tightened bounding performance as shown in Fig. \ref{Fignewupperbound}. 
 
 \begin{figure}[htbp]
 	\centering
 	\subfigure[Rate performance over the AWGN channel with $c=8$.]
 	{\includegraphics[angle=0,width=0.48\textwidth]{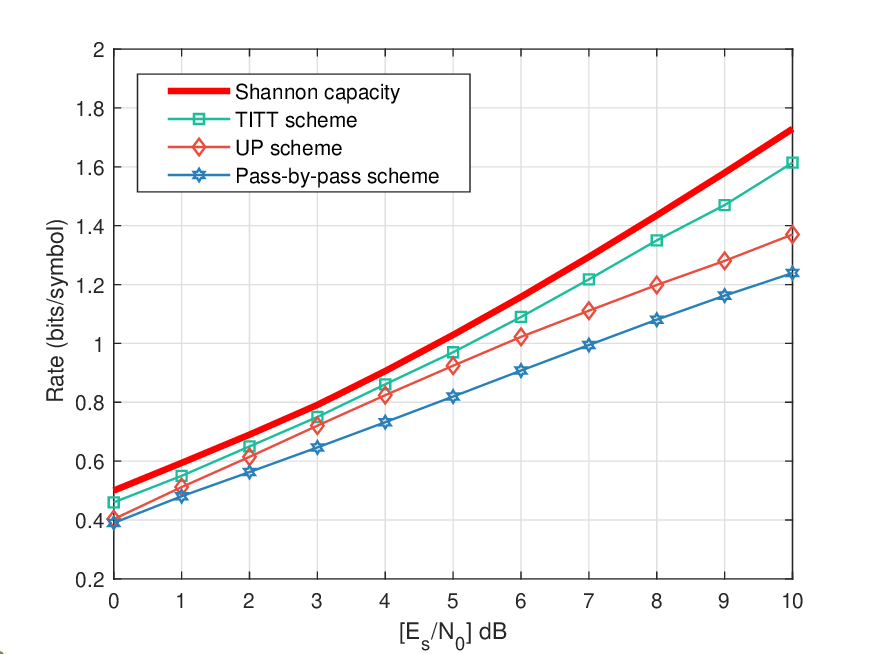}}
 	\subfigure[Rate performance over the BSC with $c=1$.]
 	{\includegraphics[angle=0,width=0.48\textwidth]{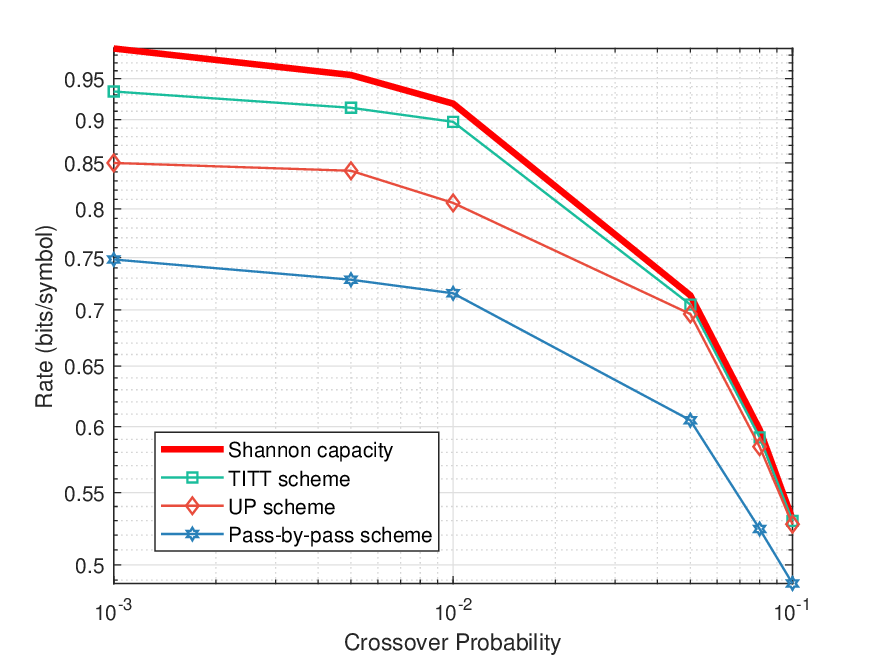}}
 	\caption{Rate performance of different transmission schemes with $n=32$, $v=32$, $k=4$ and $B=64$.}\label{f:BER performance over AWGN channel.}\label{fig5}
 \end{figure}

 Fig. \ref{FER comparision} is obtained over the BSC model, which provides the numerical comparison among the upper bound proposed in \cite{ref18}, the upper bound given in (\ref{BSCb}), and the Monte Carlo simulation results. For the simulation setup, we also select $n$ as short as $n=8$ and set the Monte Carlo Simulation sample size as $10^4$ to obtain the simulated error probability. From Fig. \ref{FER comparision}, we can see a superior bounding performance of the proposed upper bound with wide ranges of SNR. That is, the proposed upper bound remains a closer gap with the Monte Carlo Simulation results compared to that in \cite{ref18}. As the proposed bound further narrows the gap with the simulated results, it can provide a more exact performance approximation for Spinal codes.
 \subsection{Rate Performance}
 
 Fig. \ref{fig5} shows the rate performance comparisons among the proposed TITT transmission scheme, the uniform-puncturing-based transmission scheme (referred to as UP scheme), and the pass-by-pass transmission scheme. The simulation results in Fig. \ref{fig5} (a) are conducted over the AWGN channel model, while those in Fig. \ref{fig5} (b) are obtained over the BSC model. Here we leverage the bubble decoding algorithm with $B=64$ for the simulation setup. It can be seen that the proposed TITT scheme outperforms the UP scheme and the pass-by-pass transmission scheme over both the BSC and the AWGN channel. 

%

%

\subsection{Decoding Complexity}

\begin{figure}
	\centering
	\includegraphics[angle=0,width=0.8\textwidth]{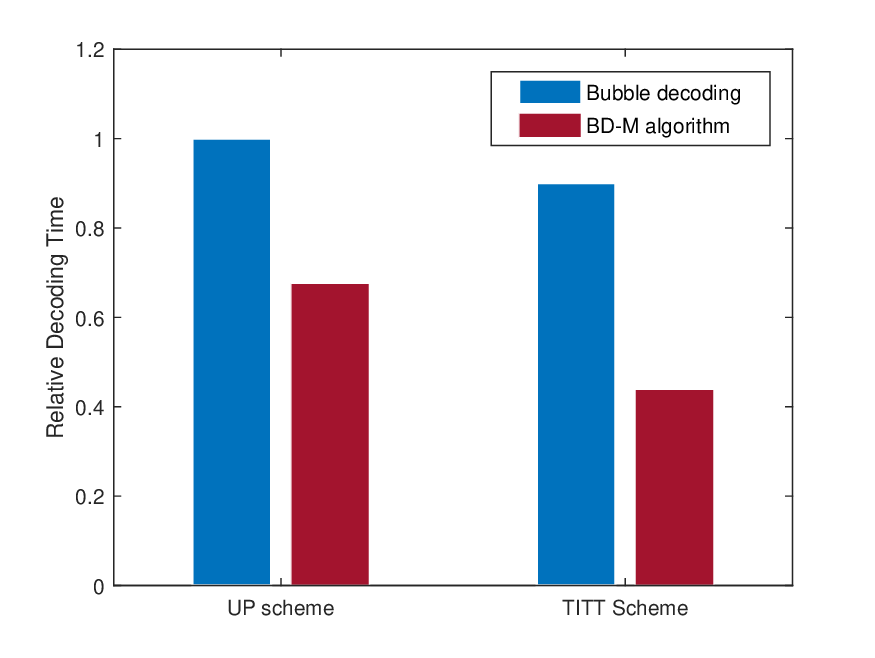}
	\caption{The average normalized decoding time of different `transmission scheme - decoding algorithm' pairs. }\label{f:puncturing strategy over AWGN channel}
\end{figure}
Fig. \ref{f:puncturing strategy over AWGN channel} gives the average normalized decoding time of different `transmission scheme - decoding algorithm' pairs. For the comparison purpose, we count the elapsed time required for successfully decoding $10^4$ packets and obtain the relative decoding time as in Fig. \ref{f:puncturing strategy over AWGN channel}. From Fig. \ref{f:puncturing strategy over AWGN channel}, we can observe that the proposed TITT scheme exhibits lower complexity than the uniform-puncturing-based transmission scheme. The reason is that the TITT scheme requires fewer coded symbols for successful decoding, thereby reducing the decoding attempts at the decoder. 

Furthermore, we can also observe that the proposed BD-M algorithm reduces the decoding time for both the UP scheme and the proposed TITT scheme, demonstrating its superiority of decoding efficiency compared to the original bubble decoding algorithm. In comparison, the reduced decoding time for TITT scheme is more evident than that for the UP, verifying the compatibility between the proposed BD-M algorithm and the TITT transmission scheme. Combining the simulation results given in Fig. \ref{fig5} and Fig. \ref{f:puncturing strategy over AWGN channel}, we find that the designed TITT scheme at the transmitter and the proposed BD-M algorithm at the receiver jointly constitute a dynamic transmission-decoding system for Spinal codes. With this combination implementation, both the rate performance and the decoding complexity of Spinal codes are improved.

\section{Conclusions}
\label{section7}
This paper analyzed the error probability of Spinal codes over both the BSC and the AWGN channel and derived new upper bounds on the ML decoding error probability. Based on the derived upper bounds, the improved transmission scheme is designed to increase the code rate of Spinal codes further. Furthermore, to address the issue of decoding complexity, we prove that the decoding tree only changes partially during the dynamic transmission-decoding process. Inspired by this idea, we
design a partial-tree-updating BD-M algorithm, which matches the TITT scheme perfectly. 

The work in this paper may also stimulate further research interests and efforts in related topics. The derived upper bounds can provide theoretical support and guidance for designing other high-efficiency coding-associated techniques, such as unequal error protection and concatenation with outer codes. In addition, the idea of partial-tree-updating decoding also provides a framework to further improve the decoding efficiency of other state-of-the-art decoding algorithms in the literature. For example, the FSD with memory and the SFD with memory algorithms may also be invented.


%

\appendices
\section{An approach to calculate $E_0\left({\mathcal{Q}}\right)$}
For our considered Spinal codes based system, the channel input set is $\Omega = \left\{0,1,2,\cdots,2^c-1\right\}$, wherein each coded symbol follows the uniform distribution with $\mathcal{Q}\left(x\right)=2^{-c}$. Applying these in (\ref{E0Q}) results in the equality: 

\begin{equation}
\nonumber
\begin{aligned}
{E_0}\left( \mathcal{Q} \right) &=  - \log \left\{ {\frac{1}{{\sqrt {2\pi {\sigma ^2}} }} \times \int_\mathbb{R} {{{\left( {\sum\limits_{i \in \Omega } {{2^{ - c}} \cdot \exp \left( { - \frac{{{{\left( {y - i} \right)}^2}}}{{4{\sigma ^2}}}} \right)} } \right)}^2}dy} } \right\} \\ 
&=  - \log \left\{ {\frac{{{2^{ - 2c}}}}{{\sqrt {2\pi {\sigma ^2}} }} \times \int_\mathbb{R} {{{\left( {\sum\limits_{i \in \Omega } {\exp \left( { - \frac{{{{\left( {y - i} \right)}^2}}}{{4{\sigma ^2}}}} \right)} } \right)}^2}dy} } \right\} \\ 
&=  - \log \left\{ {\frac{{{2^{ - 2c}}}}{{\sqrt {2\pi {\sigma ^2}} }} \times \int_\mathbb{R} {\left( {\sum\limits_{j \in \Omega } {\sum\limits_{i \in \Omega } {\exp \left( { - \frac{{{{\left( {y - i} \right)}^2}}}{{4{\sigma ^2}}} - \frac{{{{\left( {y - j} \right)}^2}}}{{4{\sigma ^2}}}} \right)} } } \right)dy} } \right\} \\ 
&=  - \log \left\{ {\frac{{{2^{ - 2c}}}}{{\sqrt {2\pi {\sigma ^2}} }} \times \sum\limits_{j \in \Omega } {\sum\limits_{i \in \Omega } {\int_\mathbb{R} {\exp \left( { - \frac{{{{\left( {y - i} \right)}^2}}}{{4{\sigma ^2}}} - \frac{{{{\left( {y - j} \right)}^2}}}{{4{\sigma ^2}}}} \right)dy} } } } \right\} \\ 
&=  - \log \left\{ {\frac{{{2^{ - 2c}}}}{{\sqrt {2\pi {\sigma ^2}} }} \times \sum\limits_{j \in \Omega } {\sum\limits_{i \in \Omega } {\int_\mathbb{R} {\exp \left( { - \frac{{2{{(y - \frac{{i + j}}{2})}^2} + \frac{1}{2}{{\left( {i - j} \right)}^2}}}{{4{\sigma ^2}}}} \right)dy} } } } \right\} \\ 
&=  - \log \left\{ {\frac{{{2^{ - 2c}}}}{{\sqrt {2\pi {\sigma ^2}} }} \times \sum\limits_{j \in \Omega } {\sum\limits_{i \in \Omega } {\exp \left( { - \frac{{{{\left( {i - j} \right)}^2}}}{{8{\sigma ^2}}}} \right)\int_\mathbb{R} {\exp \left( { - \frac{{{{\left( {y - \frac{{i + j}}{2}} \right)}^2}}}{{2{\sigma ^2}}}} \right)dy} } } } \right\} \\ 
&=  - \log \left\{ {{2^{ - 2c}} \times \sum\limits_{j \in \Omega } {\sum\limits_{i \in \Omega } {\exp \left( { - \frac{{{{\left( {i - j} \right)}^2}}}{{8{\sigma ^2}}}} \right)} } } \right\}.
\end{aligned}
\end{equation}

By calculating $\sum\limits_{j \in \Omega } {\sum\limits_{i \in \Omega } {\exp \left( { - \frac{{{{\left( {i - j} \right)}^2}}}{{8{\sigma ^2}}}} \right)} }$, one can obtain the numerical result of $E_0\left(\mathcal{Q}\right)$.  
\section{Proof of Lemma 1}
We notice that if we let $r=\sigma x$ and compute the differential $dr=\sigma dx$, we can obtain that
\begin{equation}\label{substitution}
\int\limits_{0}^{\mathcal{D}} {{r^{2{ L_a} - 1}}} {e^{ - \frac{{{r^2}}}{{2{\sigma ^2}}}}}dr\overset{r=\sigma x}{=}\sigma^{2L_a}\cdot\int\limits_{0}^{\mathcal{D/\sigma}} {{x^{2{ L_a} - 1}}} {e^{ - \frac{{{x^2}}}{{2}}}}dx.
\end{equation}

We begin with considering the indefinite form of the RHS of (\ref{substitution}). Note that the indefinite integral ${\int {{x^{2{ L_a} - 1}}} {e^{ - \frac{{{x^2}}}{{2}}}}dr}$ is composed by two parts: $x^{2L_a-1}$ and $e^{\frac{-x^2}{2\sigma^2}}$, we can do this integral by iteratively introducing the method of integration by parts as:

\begin{equation}
\nonumber
\begin{array}{c}
{\int {{x^{2{ L_a} - 1}}} {e^{ - \frac{{{x^2}}}{{2}}}}dx} =  { - {x^{2L_a - 2}}{e^{ - \frac{{{x^2}}}{2}}}}  + \left( {2L_a - 2} \right)\int {{x^{2L_a - 3}}} {e^{ - \frac{{{x^2}}}{2}}}dx,\\
{\int {{x^{2{ L_a} - 3}}} {e^{ - \frac{{{x^2}}}{{2}}}}dx} =  { - {x^{2L_a - 4}}{e^{ - \frac{{{x^2}}}{2}}}}  + \left( {2L_a - 4} \right)\int {{x^{2L_a - 5}}} {e^{ - \frac{{{x^2}}}{2}}}dx,\\
\vdots \\
{\int {{x^{3}}} {e^{ - \frac{{{x^2}}}{{2}}}}dx} =  { - {x^{2}}{e^{ - \frac{{{x^2}}}{2}}}}  + 2\int {{x}} {e^{ - \frac{{{x^2}}}{2}}}dx,\\
{\int {{x}} {e^{ - \frac{{{x^2}}}{{2}}}}dx} =  -e^{ - \frac{{{x^2}}}{2}} +C .
\end{array}
\end{equation} 
Hence, let $a_{n}\left(x\right)=\int x^n e^{-\frac{x^2}{2}}dx$ and we obtain the recurrence relation that 
\begin{equation}
\nonumber
\begin{array}{c}
a_{2n-1}\left(x\right)=-x^{2n-2}e^{-\frac{x^2}{2}}+\left(2n-2\right)a_{2n-3}\left(x\right), n \in \mathbb{N}^+\\
a_1(x)= -e^{ - \frac{{{x^2}}}{2}} +C.
\end{array}
\end{equation}
With the recursive formula we obtain the explicit expression of $a_{2n-1}\left(x\right)$ as:
\begin{equation} \label{a2n-1}
a_{2n-1}\left(x\right)={ - {e^{\frac{{ - {x^2}}}{2}}}\sum\limits_{i = 1}^n {{x^{2\left( {n - i} \right)}}} \prod\limits_{j = 1}^{i - 1} {2\left( {n - j} \right)} } +C_1.
\end{equation}
Applying (\ref{a2n-1}) in (\ref{substitution}) results in the solution that
\begin{equation}\label{2la-1}
\begin{aligned}
&\int\limits_{0}^{\mathcal{D}} {{r^{2{ L_a} - 1}}} {e^{ - \frac{{{r^2}}}{{2{\sigma ^2}}}}}dr=\sigma^{2L_a}\left(a_{2L_a-1}\left({\mathcal{D}}/{\sigma}\right)-a_{2L_a-1}\left(0\right)\right)\\
&={\sigma ^{2 L_a}}\left( { - {e^{\frac{{ - { \mathcal{D}^2}}}{{2{\sigma ^2}}}}}\sum\limits_{i = 1}^ {L_a} {\frac{{{ \mathcal{D}^{2\left( { L_a - i} \right)}}}}{{{\sigma ^{2\left( { L_a - i} \right)}}}}} \prod\limits_{j = 1}^{i - 1}2\left( {n - j} \right) + \prod\limits_{j = 1}^{L_a - 1}2\left( {n - j} \right)  } \right).
\end{aligned}
\end{equation}
Let $\mathcal{I}_i\triangleq\prod\limits_{j = 1}^{i} {2\left( { L_a - j} \right)}$ and we can simply (\ref{2la-1}) as 
\begin{equation}
{\int\limits_0^\mathcal{D} {{r^{2{ L_a} - 1}}} {e^{ - \frac{{{r^2}}}{{2{\sigma ^2}}}}}dr}={\sigma ^{2 L_a}}\left( { - {e^{\frac{{ - { \mathcal{D}^2}}}{{2{\sigma ^2}}}}}\sum\limits_{i = 1}^ {L_a} {\frac{{{ \mathcal{D}^{2\left( { L_a - i} \right)}}}}{{{\sigma ^{2\left( { L_a - i} \right)}}}}} \mathcal{I}_{i-1} + \mathcal{I}_{L_a-1}  } \right).
\end{equation}
\section{Proof of Lemma 2}
Let $r=\sigma x$ and we obtain that
\begin{equation} \label{substitution2}
\int\limits_{\mathcal{D}}^{\infty} {{r^{{ L_a} - 1}}} {e^{ - \frac{{{r^2}}}{{2{\sigma ^2}}}}}dr\overset{r=\sigma x}{=}\sigma^{L_a}\cdot\int\limits_{\mathcal{D}/\sigma}^{\infty} {{x^{{ L_a} - 1}}} {e^{ - \frac{{{x^2}}}{{2}}}}dx.
\end{equation}

If $L_a$ is even, the solution to indefinite integral $\int x^{L_a-1} e^{-\frac{x^2}{2}}dx$ can be obtained from (\ref{a2n-1}): 
\begin{equation}
\begin{aligned}
&\int x^{L_a-1} e^{-\frac{x^2}{2}}dx = a_{L_a-1}\left(x\right)\overset{L_a=2n}{=}{a_{2n-1}\left(x\right)}\\
&={ - \sum\limits_{i = 1}^n {{x^{2\left( {n - i} \right)}}} \prod\limits_{j = 1}^{i - 1} {2\left( {n - j} \right){e^{\frac{{ - {x^2}}}{2}}}} } +C_1\\
&\overset{n=\frac{L_a}{2}}{=}{ - \sum\limits_{i = 1}^{L_a/2 }{{x^{2\left( {L_a/2 - i} \right)}}} \prod\limits_{j = 1}^{i - 1} {2\left( {L_a/2 - j} \right){e^{\frac{{ - {x^2}}}{2}}}} } +C_1.
\end{aligned}
\end{equation}
Let $\mathcal{K}_i\triangleq\prod\limits_{j = 1}^{i} {\left( { L_a - 2j} \right)}$, and the definite integral can be calculated by
\begin{equation}
\int\limits_{\mathcal{D}/\sigma}^{\infty} {{x^{{ L_a} - 1}}} {e^{ - \frac{{{x^2}}}{{2}}}}dx={e^{\frac{{ - {\mathcal{D}^2}}}{{2{\sigma ^2}}}}}\sum\limits_{i = 1}^{{L_a}/2} {\frac{{{\mathcal{D}^{\left( {{L_a} - 2i} \right)}}}}{{{\sigma ^{\left( {{L_a} - 2i} \right)}}}}} \mathcal{K}_{i-1}.
\end{equation}

If $L_a$ is odd, notice that $a_{L_a-1}\left(x\right)\overset{L_a=2n+1}{=}{a_{2n}\left(x\right)}$, we can iteratively introduce the method of integration by parts and obtain the recurrence relation that:

\begin{equation}
\nonumber
\begin{array}{c}
a_{2n}\left(x\right)=-x^{2n-1}e^{-\frac{x^2}{2}}+\left(2n-1\right)a_{2n-3}\left(x\right), n \in \mathbb{N}^+,\\
a_0(x)= {\int {e^{ - \frac{{{x^2}}}{{2}}}}dx}.
\end{array}
\end{equation}
By solving the above recursive formula, we can obtain the explicit expression of $a_{2n}\left(x\right)$ as:
\begin{equation}
a_{2n}\left(x\right)=- \sum\limits_{i = 1}^n {{x^{2\left( {n - i} \right) + 1}}} \prod\limits_{j = 1}^{i - 1} {\left( {2\left( {n - j} \right) + 1} \right){e^{\frac{{ - {x^2}}}{2}}}}  + \prod\limits_{j = 1}^{{n} - 1} {\left( {2\left( {n - j} \right) + 1} \right)a_0\left(x\right)} .
\end{equation}
As such, we have:
\begin{equation}
\begin{aligned}
&\int x^{L_a-1} e^{-\frac{x^2}{2}}dx = a_{L_a-1}\left(x\right)\overset{L_a=2n+1}{=}{a_{2n}\left(x\right)}\\
&=- {e^{\frac{{ - {x^2}}}{2}}}\sum\limits_{i = 1}^n {{x^{2\left( {n - i} \right) + 1}}} \prod\limits_{j = 1}^{i - 1} {\left( {2\left( {n - j} \right) + 1} \right)}  + \prod\limits_{j = 1}^{{n} - 1} {\left( {2\left( {n - j} \right) + 1} \right)a_0\left(x\right)}
\\
&=- {e^{\frac{{ - {x^2}}}{2}}}\sum\limits_{i = 1}^{\left(L_a-1\right)/2} {{x^{L_a-2i} }} \mathcal{K}_{i-1} + a_0\left(x\right)\mathcal{K}_{\left( L_a-3 \right)/2}.
\end{aligned}
\end{equation}
Note that
\begin{equation}
\lim\limits_{x \to \infty}{a_0\left(x\right)}-a_0\left(\mathcal{D/\sigma}\right)=\sqrt{2 \pi} Q\left(\mathcal{D/\sigma}\right),
\end{equation}
where $Q\left(\cdot\right)$ is the Q function with $Q\left(x\right)= \int_{x}^{\infty}\frac{1}{\sqrt{2 \pi}}e^{-\frac{t^2}{2}}dt$. The definite integral in the RHS of (\ref{substitution2}) is then given as:
\begin{equation} \label{bound5}
\begin{aligned}
\int\limits_{\mathcal{D}/\sigma}^{\infty} {{x^{{ L_a} - 1}}} {e^{ - \frac{{{x^2}}}{{2}}}}dx&={e^{\frac{{ - {\mathcal{D}^2}}}{{2{\sigma ^2}}}}}\sum\limits_{i = 1}^{\left( {{L_a} - 1} \right)/2} {\frac{{{\mathcal{D}^{{L_a} - 2i}}}}{{{\sigma ^{{L_a} - 2i}}}}} \mathcal{K}_{i-1}+\mathcal{K}_{\left(L_a-3\right)/2}{\int\limits_{\mathcal{D}/\sigma}^{\infty} {e^{ - \frac{{{x^2}}}{{2}}}}dx}\\
&={e^{\frac{{ - {\mathcal{D}^2}}}{{2{\sigma ^2}}}}}\sum\limits_{i = 1}^{\left( {{L_a} - 1} \right)/2} {\frac{{{\mathcal{D}^{{L_a} - 2i}}}}{{{\sigma ^{{L_a} - 2i}}}}} \mathcal{K}_{i-1}+{\sqrt{2 \pi} Q\left(\mathcal{D/\sigma}\right)}\mathcal{K}_{\left(L_a-3\right)/2}
\end{aligned}
\end{equation}

Substitute (\ref{bound5}) into (\ref{substitution2}) and we obtain the solution.

\section{Proof of Theorem 4}
	 For ML decoding over the BSC, the Euclidean distance should be replaced by the Hamming distance:
	\begin{equation} 
	\nonumber
 \hat{M}= \mathop{\arg\min}_{M'\in\left\{0,1\right\}^n}\| \textbf{y}-\textbf{x}(M')\|\\
	=\mathop{\arg\min}_{M'\in\left\{0,1\right\}^n}\sum_{i=1}^{n/k}\sum_{j=1}^{\ell_{i}}\|{y}_{i,j} \oplus \operatorname{f}({x}_{i,j}({M}'))\|.
	\end{equation}	
	If the crossover probability of the BSC is $p$, we have $y_{i,j}=\operatorname{f}({x}_{i,j}({M}))\oplus e_{i,j}$, where $e_{i,j}$ obeys the Bernoulli distribution with parameter $p$.
	
	Similarly, we can classify the candidate sequence set $\left\{0,1\right\}^n$ into categories: the correct sequence and the the wrong sequences. The cost of correct candidate sequences can be calculated as
	
	\begin{equation}
	\nonumber
	D(M)=\sum_{i=1}^{n/k}\sum_{j=1}^{\ell_{i}}\|{y}_{i,j} \oplus \operatorname{f}({x}_{i,j}({M}))\|=\sum_{i=1}^{n/k}\sum_{j=1}^{\ell_{i}}e_{i,j}.
	\end{equation}	
	
	Then, it can be shown that $D(M)$ obeys a binomial distribution with
	\begin{equation}\label{eq21}
	\begin{aligned}
	\mathbb{P}\left( {D(M) = d} \right) = \left( {\begin{array}{*{20}{c}}
		{{L_1}}\\
		d
		\end{array}} \right){p^d}{\left( {1 - p} \right)^{{L_1} - d}}.
	\end{aligned}
	\end{equation}
	
	By similarly adopting the union bound of probability, we have
	\begin{equation}\label{eq22}
	\mathbb{P}\left( {{E_1}} \right) \le \sum\limits_{M' \in \mathcal{W}_1 } {\mathbb{P}\left( {D\left( M' \right) \le D(M)} \right)} {\rm{ = }} 
	\sum\limits_{d = 0}^{{L_1}} {\mathbb{P}\left( {D(M) = d} \right)\sum\limits_{M'\in \mathcal{W}_1} {\mathbb{P}\left( {D\left( M' \right) \le d} | D(M)=d\right)} } .
	\end{equation}
	
	The cost of wrong candidate sequence $D(M')$ can be calculated as 
	\begin{equation}
	\nonumber
	D(M') =\sum_{i=1}^{n/k}\sum_{j=1}^{\ell_{i}}\|\operatorname{f}({x}_{i,j}({M})) \oplus e_{i,j} \oplus \operatorname{f}({x}_{i,j}({M'})) \|.
	\end{equation}
	
	Note that $W'\in\mathcal{W}_1$, we can verify that all the coded symbols $\operatorname{f}(x_{i,j}(M'))$ follows Bernoulli distribution with parameter $1/2$. Thus, we have
	\begin{equation}\nonumber
	\mathbb{P}\left( {D\left( M' \right) \le d} \right) = \sum\limits_{t = 0}^d {\left( {\begin{array}{*{20}{c}}
			{{L_1}}\\
			t
			\end{array}} \right)} {2^{ - {L_1}}}.
	\end{equation}
	It can be noted that the distribution of $D(M')$ is not related to the noise $\bf e$ and the correct encoded symbols generated by ${\bf x}(M)$. Then, it holds that
	\begin{equation}\label{eq23}
	{\mathbb{P}\left( {D\left( M' \right) \le d} | D(M)=d\right)} = \mathbb{P}\left( {D\left( M' \right) \le d} \right) = \sum\limits_{t = 0}^d {\left( {\begin{array}{*{20}{c}}
			{{L_1}}\\
			t
			\end{array}} \right)} {2^{ - {L_1}}}
	\end{equation}
	
	By substituting (\ref{eq21}) and (\ref{eq23}) into (\ref{eq22}), we have
	\begin{equation} \label{eq24}
	\begin{aligned}
	{\mathbb{P}\left({E_{1}}\right)} \le\sum\limits_{d = 0}^{{L_1}} {\left( {\left( {\begin{array}{*{20}{c}}
				{{L_1}}\\
				d
				\end{array}} \right){p^a}{{\left( {1 - p} \right)}^{{L_1} - d}} \cdot \min \left( {1,{R_{1,d}}} \right)} \right)},
	\end{aligned}
	\end{equation}
	where $R_{1,d}$ is calculated using
	\begin{equation} \nonumber
	{R_{1,d}} = \left( {{2^k} - 1} \right){2^{n - k}}\sum\limits_{t = 0}^d {\left( {\begin{array}{*{20}{c}}
			{{L_1}}\\
			t
			\end{array}} \right)} {2^{ - {L_1}}}.
	\end{equation}  
	
The process above can be performed similarly and it holds that
	\begin{equation}\label{eq25}
	\mathbb{P}({E}_{a} | \overline{E}_{1}, \ldots, \overline{E}_{a-1})\le\sum\limits_{d = 0}^{{\ell_i}} {\left( {\left( {\begin{array}{*{20}{c}}
				{{L_a}}\\
				d
				\end{array}} \right){p^d}{{\left( {1 - p} \right)}^{{L_a} - d}} \cdot \min \left( {1,{R_{a,d}}} \right)} \right)}.\\
	\end{equation}

	At last, by substituting (\ref{eq24}) and (\ref{eq25}) into (\ref{eq16}), the proof of Theorem 4 is finished.
\section{Proof of Theorem 5}
	Note that the cost calculation at each decoding tree layer is implemented by
\begin{equation}\label{layer1}
C_\lambda=\sum_{i=1}^{\lambda}\sum_{j=1}^{\ell_{i}}\|{y}_{i,j}-\operatorname{f}({x}_{i,j}({M}'))\|^2,
\end{equation}
	Provided that $y_{a,\ell_a+1}$ is newly received, and denote cost of the updated decoding tree at layer $\lambda$ as $C'_{\lambda}$, then the costs at layer $\lambda < a$ remains the same as (\ref{layer1}), with
\begin{equation}
C'_\lambda=\sum_{i=1}^{\lambda}\sum_{j=1}^{\ell_{i}}\|{y}_{i,j}-\operatorname{f}({x}_{i,j}({M}'))\|^2=C_{\lambda}, {\text{\rm for }\lambda < a}.
\end{equation}
	However, for $\lambda \ge a$, $C'_{\lambda}$ turns to
\begin{equation}
\begin{aligned}
C'_\lambda&=\sum_{i=1}^{\lambda}\sum_{j=1}^{\ell_{i}}\|{y}_{i,j}-\operatorname{f}({x}_{i,j}({M}'))\|^2 +\|{y}_{a,\ell_a+1}-\operatorname{f}({x}_{a,\ell_a+1}({M}'))\|^2\\
&=C_{\lambda}+\|{y}_{a,\ell_a+1}-\operatorname{f}({x}_{a,\ell_a+1}({M}'))\|^2,{\text{\rm for }\lambda \ge a}.
\end{aligned}
\end{equation}

\section{Proof of (\ref{eq20})}
For the improved transmission scheme which combines the uniform puncturing with the incremental tail transmission, it is easy to find that
\begin{equation}\label{eq26}
{\ell'_1} \le {\ell_1},{\ell'_2} \le {\ell_2}, \dots, {\ell'_{n/k-1}} \le {\ell_{n/k-1}}, {\ell'_{n/k}} \ge {\ell_{n/k}}.
\end{equation}

As the rate corresponding to the improved transmission scheme is higher than the rate when using uniform puncturing, it turns out that
 \begin{equation}\label{eq27}
\sum\limits_{i = 1}^{n/k} {{\ell_i}}  \ge \sum\limits_{i = 1}^{n/k} {{\ell'_i}}.
 \end{equation}

According to (\ref{eq26}), we assume that ${\ell'_i} + {r_i} = {\ell_i}$ for $ 1 \le i \le n/k - 1$ and ${\ell'_{n/k}} - {r_{n/k}} = {\ell_{n/k}}$, where ${r_i} \ge 0$. Substituting them into (\ref{eq27}), we can have
\begin{equation}
\sum\limits_{i = 1}^{n/k - 1} {{r_i}}  \ge {r_{n/k}}.
\end{equation}

Therefore, we obtain that

\begin{equation}
\begin{aligned}
\sum\limits_{i = 1}^{n/k} {\left( {{\ell_i} - {\ell'_i}} \right){o_i}}  &= \sum\limits_{i = 1}^{n/k - 1} {{r_i}{o_i} - {r_{n/k}}{o_{n/k}}} \\
&\ge \sum\limits_{i = 1}^{n/k - 1} {{r_i}{o_i} - \sum\limits_{i = 1}^{n/k - 1} {{r_i}} {o_{n/k}}} \\
&= \sum\limits_{i = 1}^{n/k - 1} {{r_i}\left( {{o_i} - {o_{n/k}}} \right)}  > 0.
\end{aligned}
\end{equation}


\ifCLASSOPTIONcaptionsoff
  \newpage
\fi

\end{document}